\documentclass{aa}  

\usepackage{graphicx}
\usepackage{arydshln}
\usepackage{txfonts}
\usepackage{xcolor}
\usepackage[normalem]{ulem}
\begin{document}

   \title{How primordial is the structure of comet 67P/C-G?}

   \subtitle{Combined collisional and dynamical models suggest a late formation}

   \author{M. Jutzi
          \inst{1}, W. Benz\inst{1}, A. Toliou\inst{2,3}, A. Morbidelli\inst{3}, R. Brasser\inst{4}
                    }

   \institute{Physics Institute, University of Bern, NCCR PlanetS,
              Sidlerstrasse 5, 3012 Bern, Switzerland\\
              \email{martin.jutzi@space.unibe.ch; willy.benz@space.unibe.ch}
         \and
         Department of Physics, Aristotle University of Thessaloniki, GR-54124 Thessaloniki, Greece\\
         \email{athtolio@physics.auth.gr}
         \and
            Laboratoire Lagrange, Universit\'e C\^ote d'Azur, CNRS, Observatoire de la C\^ote d'Azur, Nice, France\\
         \and
         Earth Life Science Institute, Tokyo Institute of Technology, Meguro-ku, Tokyo, 152-8550, Japan
}

   \date{Received -- ; accepted --}

 
  \abstract
   {There is an active debate about whether the properties of comets as observed today are primordial or, alternatively, if they are a result of collisional evolution or other processes.}
   {We investigate the effects of collisions on a comet with a structure like 67P/Churyumov-Gerasimenko (hereafter 67P/C-G). We develop scaling laws for the critical specific impact energies $Q_{reshape}$ required for a significant shape alteration. These are then used in simulations of the combined dynamical and collisional evolution of comets in order to study the survival probability of a primordially formed object with a shape like 67P/C-G. Although the focus of this work is on a structure of this kind, the analysis is also performed for more generic bi-lobe shapes, for which we define the critical specific energy $Q_{bil}$. The simulation outcomes are also analyzed in terms of impact heating and the evolution of the porosity.}
   {The effects of impacts on comet 67P/C-G are studied using a state-of-the-art smooth particle hydrodynamics (hereafter SPH) shock physics code. In the 3D simulations, a publicly available shape model of 67P/C-G is applied and a range of impact conditions and material properties are investigated. The resulting critical specific impact energy $Q_{reshape}$ (as well as $Q_{bil}$ for generic bi-lobe shapes) defines a minimal projectile size which is used to compute the number of shape-changing collisions in a set of  dynamical simulations. These simulations follow the dispersion of the trans-Neptunian disk during the giant planet instability, the formation of a scattered disk, and produce 87 objects that penetrate into the inner solar system with orbits consistent with the observed JFC population.  The collisional evolution before the giant planet instability is not considered here. Hence, our study is conservative in its estimation of the number of collisions.}
   {We find that in any scenario considered here, comet 67P/C-G would have experienced a significant number of shape-changing collisions, if it formed primordially. This is also the case for generic bi-lobe shapes. Our study also shows  that impact heating is very localized and that collisionally processed bodies can still have a high porosity.}
   {Our study indicates that the observed bi-lobe structure of comet 67P/C-G may not be primordial, but might have originated in a rather recent event, possibly within the last 1 Gy. This may be the case for any kilometer-sized two-component cometary nuclei.}


   \maketitle

%

\section{Introduction}

Comets or their precursors formed in the outer planet region during the initial stages of solar system formation. They may have been assembled by hierarchical accretion \citep[e.g.][]{Weidenschilling:1997im,Windmark:2012wbo,Windmark:2012wbg,Kataoka:2013kt} or, alternatively, were born big in gravitational instabilities  \citep[e.g.][]{Youdin:2005yg,Johansen:2007jo,Cuzzi:2010iv,Morbidelli:2009dd}, thereby bypassing the primary accretion phase entirely. Whether the cometary nuclei structures as observed today are pristine and preserve a record of their original accumulation, or are a result of later collisional or other processes is much debated (e.g. \citealp{Weissmann:2004wa,Mumma:1993mw,Sierks:2015sb,Rickman:2015wu,Morbidelli:2015vm,Jutzi:2015ja,Davidsson:2016ds}). The shape, density, composition, and other properties of comet 67P/Churyumov-Gerasimenko (67P/C-G) have been determined in detail by the European Space Agency's Rosetta rendezvous mission \citep[e.g.][]{Sierks:2015sb,Haessig:2015he,Capaccioni:2015cc}. Based on this data, it has been suggested that its structure is pristine and was formed in the early stages of the solar system \citep{Massironi:2015ma}, possibly by low velocity accretionary collisions \citep{Jutzi:2015ja}. What is less clear is whether or not a structure like comet 67P/C-G would have been able to survive until today. 

The collisional evolution of an object of the size of comet 67P/C-G was studied by \citet{Morbidelli:2015vm} using dynamical models of  the evolution of the early solar system. In the "standard model", as defined by the so-called Nice model \citep{Tsiganis:2005tg}, cometary nuclei, or their precursors, originated from an initial trans-planetary disk of icy planetesimals with a lifetime of a few hundred Myr. In this concept, the trans-planetary disk formed in the infant stages of the solar system beyond the original orbits of all giant planets, which were initially closer to the Sun. This disk may have given rise to both the Scattered Disk and the Oort cloud \citep{Brasser:2013dw} and thus, it may once have been the repository for all the comets observed today.  
According to  the standard assumption, the dispersal of the disk coincided with the beginning of the so-called Late Heavy Bombardment \citep{Gomes:2005gl,Morbidelli:2012ms}, and had a lifetime of about 450 Myr before it was dynamically dispersed.

As shown in  \citet{Morbidelli:2015vm}, it is clear that in this standard model, an object of the size of comet 67P/C-G would have experienced a high number of catastrophic collisions and thus could not have survived. However, it was also shown that in the (hypothetical) case that the dispersal of the disk occurred early, right after gas removal, the collisional evolution of km-size bodies ending in the Scattered Disk would have been less severe, and a fraction of these objects may have escaped all catastrophic collisions. We also note that in alternative models \citep[e.g.][]{Davidsson:2016ds}, the total number of comets is considered to be lower than previously thought. 
Therefore, the fate of cometary-sized objects appears to depend upon the details of the dynamical scenario considered. 

However, whether or not an object like comet 67P/C-G would have been able to survive until today does not only depend upon its dynamical evolution but even more so on the "strength" of the body. In other words, it is crucial to know the critical specific impact energy at which the shape and structure of such an object are altered significantly. Previous studies of the collisional evolution of  comet 67P/C-G  \citep{Morbidelli:2015vm}  used scaling laws for catastrophic disruption energies that are based on idealized spherical, solid icy bodies \citep{Benz:1999cj}, which may not represent well the properties of a highly porous cometary nuclei. It is well known that the impacts in highly porous material, given its dissipative properties, can lead to very different results compared to impacts involving solid materials (e.g. \citealp{Housen:2003bz,Jutzi:2008kp}). Furthermore, complex shapes such as the one of 67P/C-G may already be substantially affected by relatively low energy, sub-catastrophic impacts.

It has been suggested recently that rotational fission and reconfiguration may be a dominant structural evolution process for short-period comet nuclei having a two-component structure with a  volume ratio larger than $\sim$ 0.2 \citep{Hirabayashi:2016hs}. In this model, the fission-merging cycle would begin once a two-component body enters the inner solar system and significant changes in the rotation period occur. The final shape of the comet nuclei (e.g. 67P/C-G) as observed today would then be the result of the last merger in this cycle. In this context, it is important to also study the survival probability of more general two-component structures, as such structures are required for the fission-merging cycle to begin. 

In this paper, we consider both the dynamical evolution and the response to impacts of objects with a 67P/C-G-like shape as well as generic bi-lobe structures. This combined approach allows us to compute the expected number of shape-changing collisions for such objects, as well as the related survival probability and possible formation age. 

In the first part of the paper, we describe our modeling approach to study the effects of impacts on comet 67P/C-G and generic bi-lobe shapes. In section 2, we determine the specific energies $Q_{reshape}$ required to change a 67P/C-G-like shape significantly, as well as the corresponding $Q_{bil}$ for reshaping generic bi-lobe objects. The catastrophic disruption threshold $Q^*_D$ for bodies of 67P/C-G size, with the same properties, is computed as well here. Using the result of this modeling, we develop scaling laws for $Q_{reshape}$, $Q_{bil}$ and  $Q^*_D$. Finally, the simulation outcomes are analyzed in terms of impact heating and the evolution of the porosity.

In the second part of the paper, we first describe the details of the dynamical simulations used in this study and discuss the differences and the improvements with respect to the previous work by \citet{Morbidelli:2015vm}  (section \ref{sec:dynmodel}). We then compute the average number of shape-changing collisions for a body with a 67P/C-G-like shape as well as for generic bi-lobe shapes, using the corresponding  scaling laws ($Q_{reshape}$ and $Q_{bil}$). 
In section \ref{sec:validity}, the uncertainties of our model as well as alternative models are discussed, followed by conclusions in section  \ref{sec:conclusions}. 

A scenario of the late formation of  67P/C-G-like (two-lobe) shapes by a new type of sub-catastrophic impacts is presented in a companion paper (Jutzi\&Benz, 2016 submitted; hereafter Paper II). 


\section{The effects of impacts on bi-lobe structures}\label{sec:qshape}
Here, in a suite of 3D smooth particle hydrodynamics (SPH) code calculations, we compute the specific impact energy $Q_{reshape}$ required to significantly change the shape of comet 67P/C-G as well as of generic bi-lobe structures. The catastrophic disruption threshold $Q^*_D$ for spherical objects of the same mass is computed as well. We consider a range of material (strength) properties and various impact conditions.  The simulation outcomes are also analysed in terms of impact heating and the evolution of the porosity.

\subsection{Assumptions}

Cometary nuclei come apart easily due to tides \citep{Asphaug:1994cya} and other gentle stresses \citep{Boehnhardt:2004uu}. Laboratory materials analysis \citep{Skorov:2012ks}, observations of comet disruptions by tides \citep{Asphaug:1994cya} or fragmentation through dynamic sublimation pressure \citep{Steckloff:2015sj}, suggest a \emph{bulk} strength of $<$ 10 - 100 Pa for these weakly consolidated bodies. On the other hand, a high compressive strength of surface layers on comet 67P/C-G \citep{biele:2015bu} was found at 0.1-1 m scales. For our analysis of the overall stability, this kind of small scale (< $\sim$ 10 m) strength is not relevant, as we are interested in the bulk properties. In our modeling, we thus consider \emph{bulk} tensile strengths of up to 1 kPa. The corresponding values of cohesion and compressive strength are $\sim$ an order of magnitude higher (see section \ref{sec:impmodel}).

The low bulk densities of comets indicate substantial porosity; in the case of comet 67P/C-G it is about 75\% \citep[e.g.][]{Paetzold:2016pa}. In our modeling approach (section \ref{sec:impmodel}) it is implicitly assumed that porosity is at small scales and the body is homogenous. In the case of comet 67P/C-G, recent gravity field analysis \citep{Paetzold:2016pa} indicate that the interior of the nucleus is homogeneous (down to scales of $\sim$ 3 m) and constant in density on a global scale without large voids. This suggests our approach of modeling a homogenous interior is justified. 

   \begin{figure}[h!]
   \centering
   \includegraphics[width=1.0\hsize]{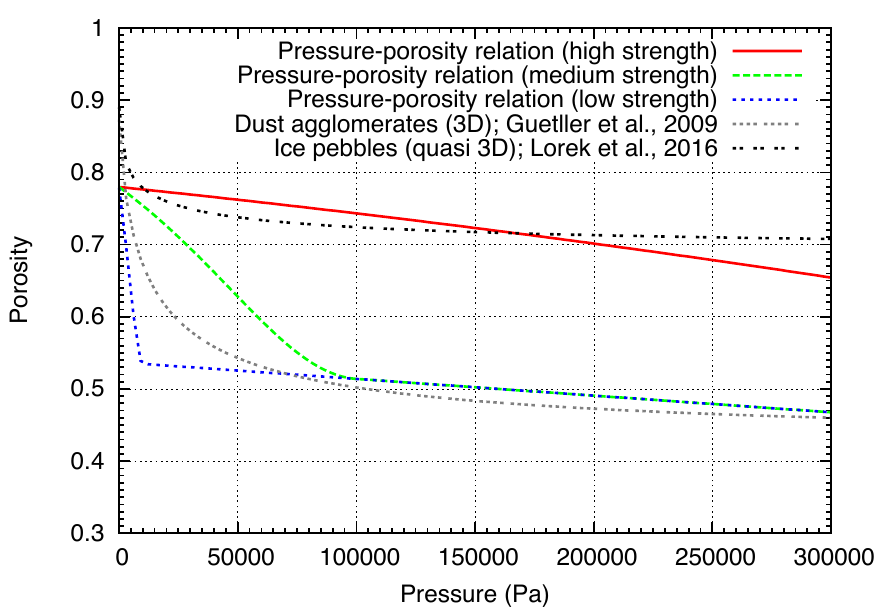}
   \caption{Pressure-porosity relations (crush curve) used in the simulations for the three different sets of parameters (low, medium, high strength) as defined in Table \ref{table:matparam}. Also shown are the results from laboratory experiments dust agglomerates \citep{Guettler:2009gk} and ice pebbles \citep{Lorek:2016lg}.}
   \label{fig:crushcurve}
    \end{figure}

\begin{table*}[ht!]
\caption{Material parameters used in the simulations. Crush curve parameters $P_e$ and $P_s$  \citep{,Jutzi:2008kp}, density of matrix material $\rho_{s0}$, initial bulk density $\rho_{0}$, density of the compacted material $\rho_{compact}$, initial distention $\alpha_{0}$, bulk modulus $A$, friction coefficient $\mu$, cohesion $Y_0$, average tensile strength $Y_T$.}            
\label{table:matparam}      
\centering                        
\begin{tabular}{l c c c c c c c c c c}        
\hline\hline            
type & $P_e$ (Pa) & $P_s$ (Pa) & $\rho_{s0}$ (kg/m$^3$) & $\rho_{0}$ (kg/m$^3$) & $\rho_{compact}$ (kg/m$^3$) & $\alpha_{0}$ & $A$ (Pa)& $\mu$	 & $Y_0$ (Pa) & $Y_T$ (Pa) \\    
\hline                        
 low strength & $10^2$ & $10^4$ & 910 & 440 & 1980 & 4.5 & 2.67$\times10^6$ & 0.55 & $10^2$ & $10^1$ \\
 medium str. (nominal) & $10^3$ & $10^5$ & 910 & 440  & 1980 & 4.5 & 2.67$\times10^6$ & 0.55 & $10^3$ & $10^2$ \\
  high strength & $10^4$ & $10^6$ & 910 & 440 & 1980 & 4.5 & 2.67$\times10^6$ & 0.55 & $10^4$ & $10^3$ \\

   \hline                               
\end{tabular}
\end{table*}

\subsection{Modeling approach}\label{sec:impmodel}
The modeling approach used here has already been successfully applied in a previous study to the regime of cometesimal collisions \citep{Jutzi:2015ja}. We use a parallel smooth particle hydrodynamics (SPH) impact code \citep{Benz:1995hx,Nyffeler:2004tz,Jutzi:2008kp,Jutzi:2015gb} which includes self-gravity as well as material strength models. To avoid numerical rotational instabilities, the scheme suggested by \citet{Speith:2006} is used.

In our modeling, we include an initial cohesion $Y_0$ $>$ 0 and use a tensile fracture model \citep{Benz:1995hx}, using a range of parameters that lead to an average tensile strength $Y_T$ varying from $\sim$ 10 to $\sim$ 1000 Pa. We consider $Y_T$ = 100 Pa as the nominal case.  To model fractured, granular material, a pressure dependent shear strength (friction) is included by using a standard Drucker-Prager yield criterion \citep{Jutzi:2015gb}. As shown in \citet{Jutzi:2015gb} and \citet{Jutzi:2015ux}, granular flow problems are well reproduced using this method. 

Porosity is modeled using a P-alpha model \citep{,Jutzi:2008kp} with a simple quadratic crush curve defined by the parameters $P_e$, $P_s$, $\rho_{0}$, $\rho_{s0}$ and $\alpha_{0}$. We further introduce the density of the compacted material $\rho_{compact}$ = 1980 kg/m$^3$, which is used define the initial distention  $\alpha_{0}=\rho_{compact}/\rho_0=4.5$ corresponding to an initial porosity of $1-1/\alpha_{0} \sim 78\%$, consistent with observations \citep{Sierks:2015sb,Kofman:2015kh,Paetzold:2016pa}. (We note that $\rho_{s0}$ in this model is a parameter determining the form of the crush curve and does not correspond to the density of the fully compacted material).  As an estimate of the compressive strength $\sigma_c$ = $P_s$/2 is used. As shown in Figure \ref{fig:crushcurve}, the pressure-porosity relations resulting from these parameters (for low, medium and high strength; Table \ref{table:matparam}) covers very well the range of experimental curves for dust agglomerates  \citep{Guettler:2009gk} and ice pebbles \citep{Lorek:2016lg}.

We apply a modified Tillotson equation of state (EOS; e.g. \citealp{Melosh:1989}) with parameters for water ice. It is adequate for modeling the collisions considered here, where the most important response is the solid compressibility. As long as there is porosity, the compressibility is limited not by the EOS but by the crush curve of the P-alpha model. The elastic wave speed $c_e$ for a porous aggregate body can be very low, of the order of $c_e$$\sim$ 0.1 km/s. To take this into account, we apply a reduced bulk modulus (leading term in the Tillotson EOS; see Table 1). The approach has the additional major advantage that the time-steps become large enough to carry out the simulations over many dynamical timescales. Different values of  $c_e$ = 10 - 100 m/s are investigated. 
 
The relevant material parameters used in the simulations are indicated in Table \ref{table:matparam}.

\subsection{Setup and initial conditions} 
\subsubsection{Impacts on comet 67P/C-G and generic bi-lobe shapes}
To setup the target, we apply a publicly available shape model of comet 67P/C-G\footnote{http://sci.esa.int/rosetta/54728-shape-model-of-comet-67p/}, which defines the surface of the body. Three different sets of material parameters as indicated in Table \ref{table:matparam} are used, corresponding to different target strength.

To determine $Q_{reshape}$ for 67P/C-G-like shapes, we investigate a range of impact energies using a range of impactor sizes of $R_p$ = 100-300 m and varying the impact velocities.  Target and impactor both have the same initial material properties; their initial bulk density is $\rho_0\sim$ 440 kg/m$^3$. 
We only consider impacts into the smaller of the the lobes of comet 67P/C-G. Two different impact geometries are investigated (Figure \ref{fig:qvargeo}). 

To determine the critical shape-changing impact energy $Q_{bil}$ in the case of more general bi-lobe structures, we set up a target consisting of two overlapping ellipsoids (Figure \ref{fig:qbil}). Each ellipsoid has an axis ratio of 0.6. The volume ratio between the two components is $\sim 0.5$ and the total mass of the body is $M_t$ = 1$\times$ $10^{13}$ kg. For these targets, we only use the nominal set of strength parameters (Table  \ref{table:matparam}) and an impactor size of $R_p$ = 100 m.

The simulations are carried out using a moderately high resolution of $\sim$ 3$\times$$10^5$ SPH particles.

\subsubsection{Collisions among spherical bodies}
In addition to  $Q_{reshape}$ and $Q_{bil}$, we also investigate the critical specific energy for catastrophic disruption $Q^*_D$ of spherical bodies of the same mass and material properties as in the model of comet 67P/C-G. To do this we consider 3 different size ratios of projectile and target  (1:2; 1:4; 1:8) and varying impact velocities. The impact angle is fixed to 45$^{\circ}$.

\subsection{Results}\label{sec:impresults}
\subsubsection{Critical specific energy for shape change}\label{sec:shaperesults}
\begin{figure*}
   \centering
   \includegraphics[width=\hsize]{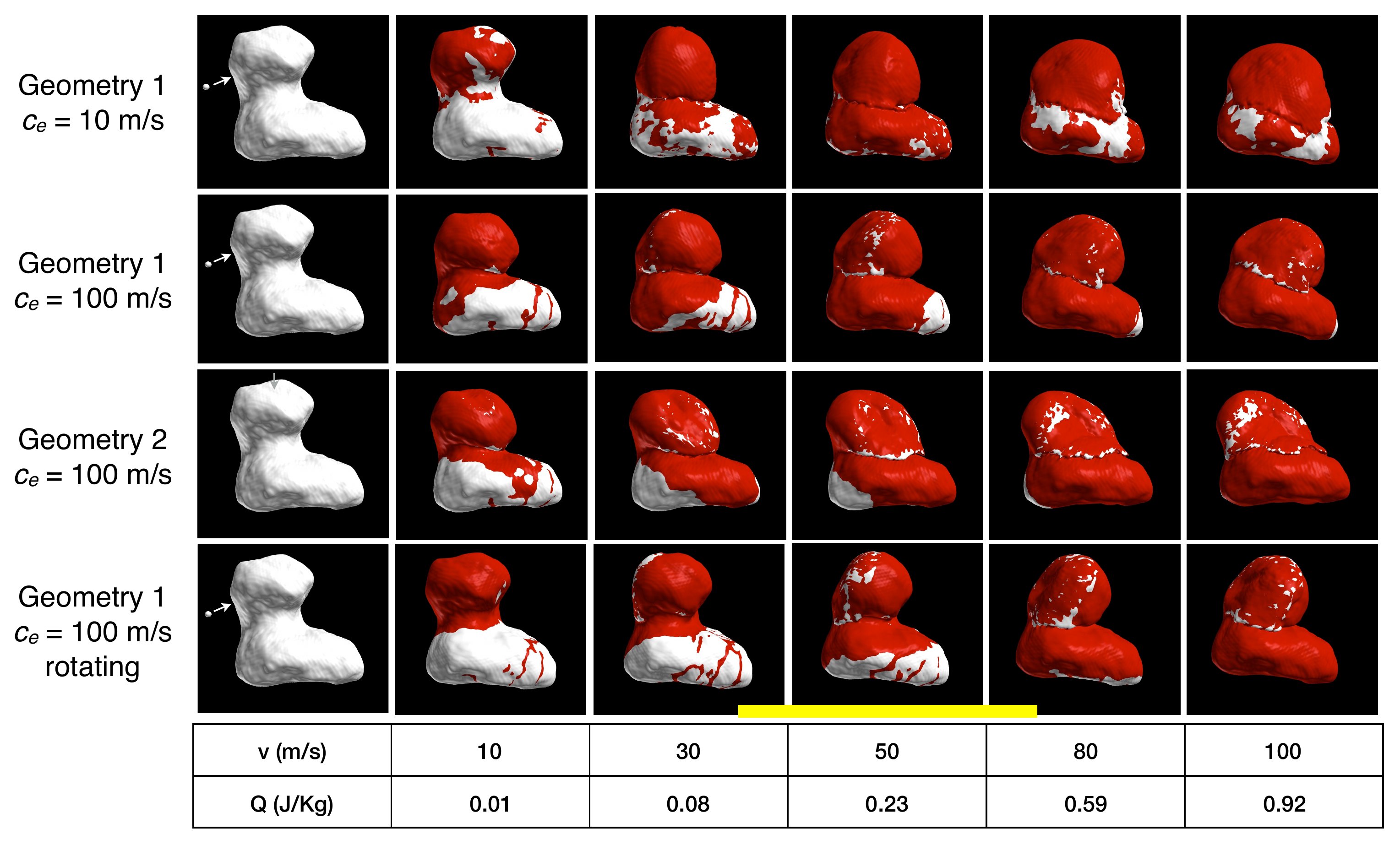}
   \caption{Shape-changing collisions on comet 67P/C-G. We use SPH to simulate impacts of a $R_p$ = 100 m projectile on the smaller of the two lobes of comet 67P/C-G. The minimal specific energy required to cause a significant change of the comet's shape by such impacts, $Q_{reshape}$, is estimated for different impact geometries and rotation axis. The material strength is the same in all cases shown here ($Y_T$ = 10 Pa). The effect of the material's sound speed is investigated as well (top row; in this case, a bulk modulus of $A$ = 2.67$\times10^4$ Pa instead of the nominal $A$ = 2.67$\times10^6$ Pa is used). Plotted is a surface of constant density which represents the surface of the comet; shown in red are regions on the surface with materials whose prescribed tensile strength was exceeded.  As a rough average, the minimal specific energy required to cause a significant shape change is estimated as $Q_{reshape} \sim$ 0.2$\pm$0.1 J/kg, as indicated by the horizontal yellow line.}
     \label{fig:qvargeo}
    \end{figure*}

\begin{figure*}
   \centering
   \includegraphics[width=\hsize]{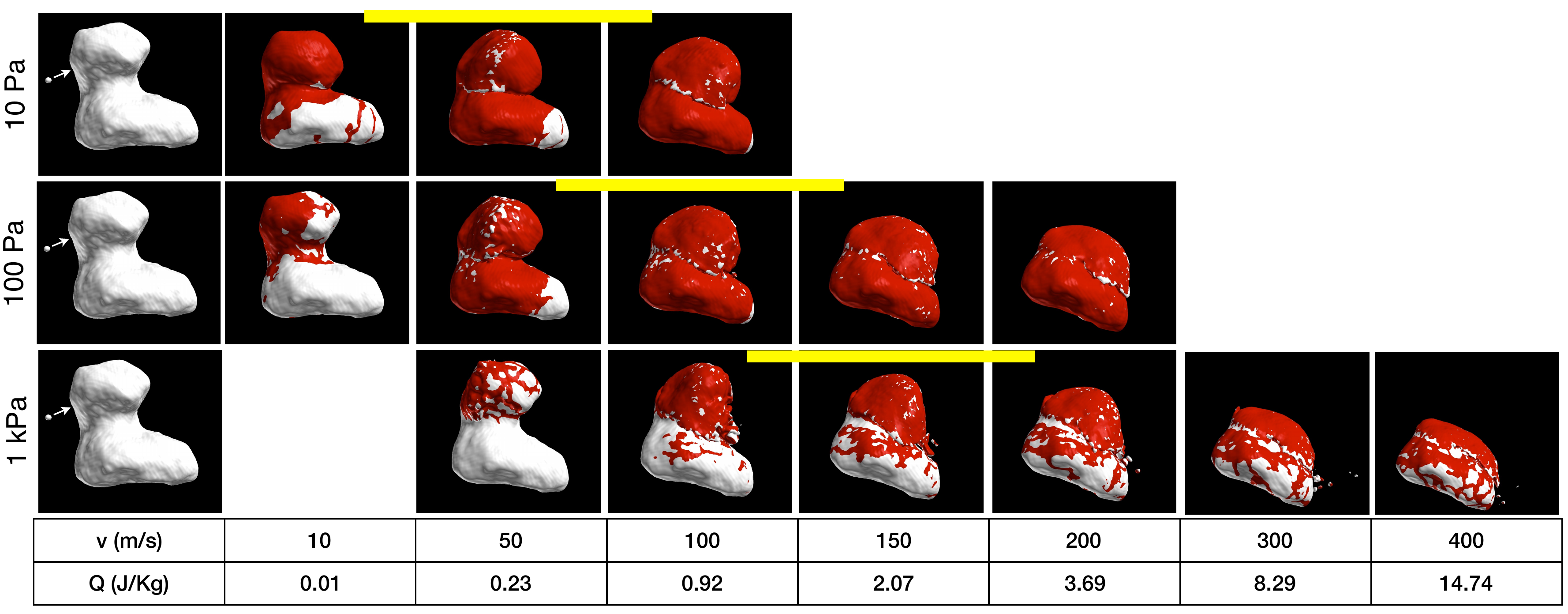}
   \caption{Same as Fig. \ref{fig:qvargeo} but for different material strength $Y_T$ of the target. $c_e$ = 100 m/s in all cases. The critical specific energies are: $Q_{reshape} \sim$ 0.2$\pm$0.1 J/kg  for $Y_T$ = 10 Pa (corresponds to average in Figure \ref{fig:qvargeo}); $Q_{reshape} \sim$ 1.0$\pm$0.5 J/kg  for $Y_T$ = 100 Pa; $Q_{reshape} \sim$ 2.0$\pm$1.0 J/kg  for $Y_T$ = 1000 Pa.}
   \label{fig:qvars}
    \end{figure*}

\begin{figure}
   \centering
   \includegraphics[width=\hsize]{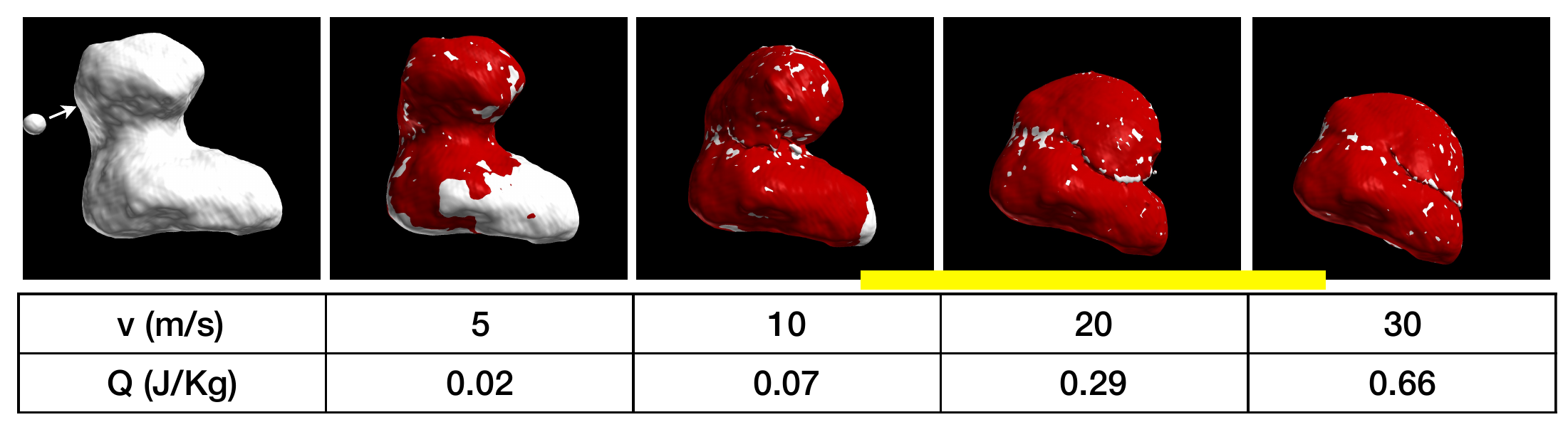}
   \caption{Same as Fig. \ref{fig:qvars} but for $R_p$ = 200 m  ($Y_T$ = 100 Pa). $Q_{reshape} \sim$ 0.3$\pm$0.15 J/kg. }
   \label{fig:qvars200}
    \end{figure}

\begin{figure}
   \centering
   \includegraphics[width=\hsize]{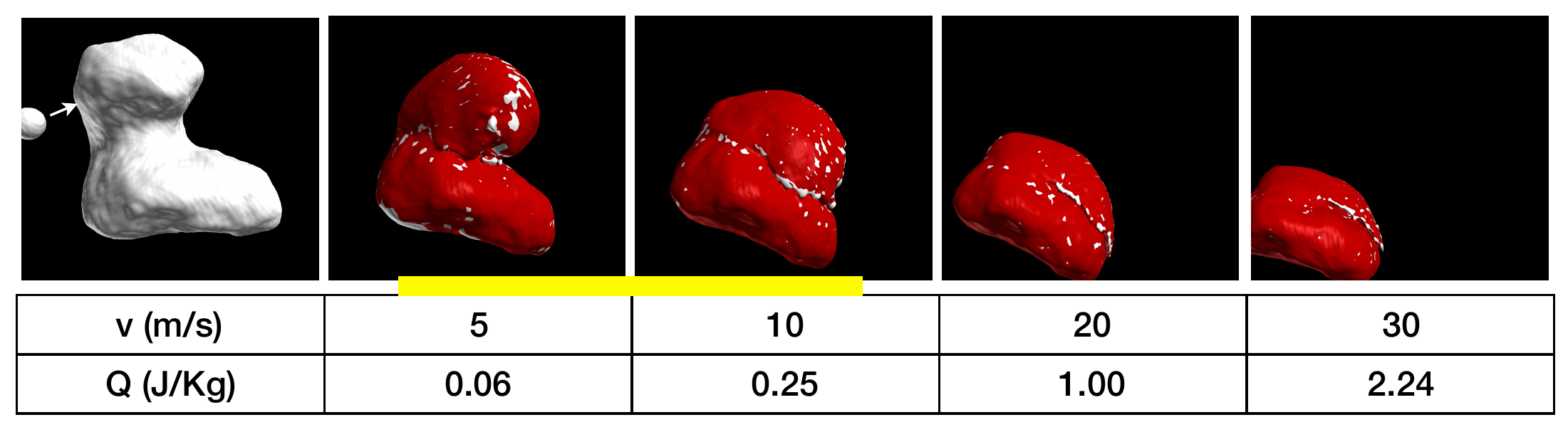}
   \caption{Same as Fig. \ref{fig:qvars} but for $R_p$ = 300 m ($Y_T$ = 100 Pa). $Q_{reshape} \sim$ 0.15$\pm$0.075 J/kg. }
   \label{fig:qvars300}
    \end{figure}

The results of our modeling of impacts on 67P/C-G are displayed in Figures \ref{fig:qvargeo}-\ref{fig:qvars300}. We find that this particular structure, with two
lobes connected by a neck, is significantly altered even by relatively low energy impacts. For a fixed set of material parameters (i.e. constant strength), the different impact geometries and rotation states considered here lead to slightly different outcomes (Figure \ref{fig:qvargeo}), but there are no major, order of magnitude, differences between the various runs. 

As it can be observed in Figure \ref{fig:qvars}, higher material strength lead to higher specific impact energy required to reach a certain degree of change in the overall shape.

There is no unique way to define the critical shape-changing specific impact energy from these results, but rough estimates are possible. Based on visual inspection, we define $Q_{reshape}$ for the different strength as: $Q_{reshape} \sim$ 0.2$\pm$0.1 J/kg  for $Y_T$ = 10 Pa; $Q_{reshape} \sim$ 1.0$\pm$0.5 J/kg  for $Y_T$ = 100 Pa; $Q_{reshape} \sim$ 2.0$\pm$1.0 J/kg  for $Y_T$ = 1000 Pa (Figure \ref{fig:qvars}) for the impacts with the  $R_p$ = 100 m projectile. For the simulations with the larger projectiles we obtain $Q_{reshape} \sim$ 0.3$\pm$0.15 J/kg  ($R_p$ = 200 m; Figure \ref{fig:qvars200}) and $Q_{reshape} \sim$ 0.15$\pm$0.075 J/kg  ($R_p$ = 300 m; Figure \ref{fig:qvars300}), using the nominal strength of $Y_T$ = 100 Pa.  These values are plotted in Figure \ref{fig:q_v_disr} and compared to the catastrophic disruption threshold, as discussed below. We note that impacts into the larger lobe may lead to slightly different values for $Q_{reshape}$, but we do not expect order of magnitude differences.

\begin{figure}
   \centering
   \includegraphics[width=\hsize]{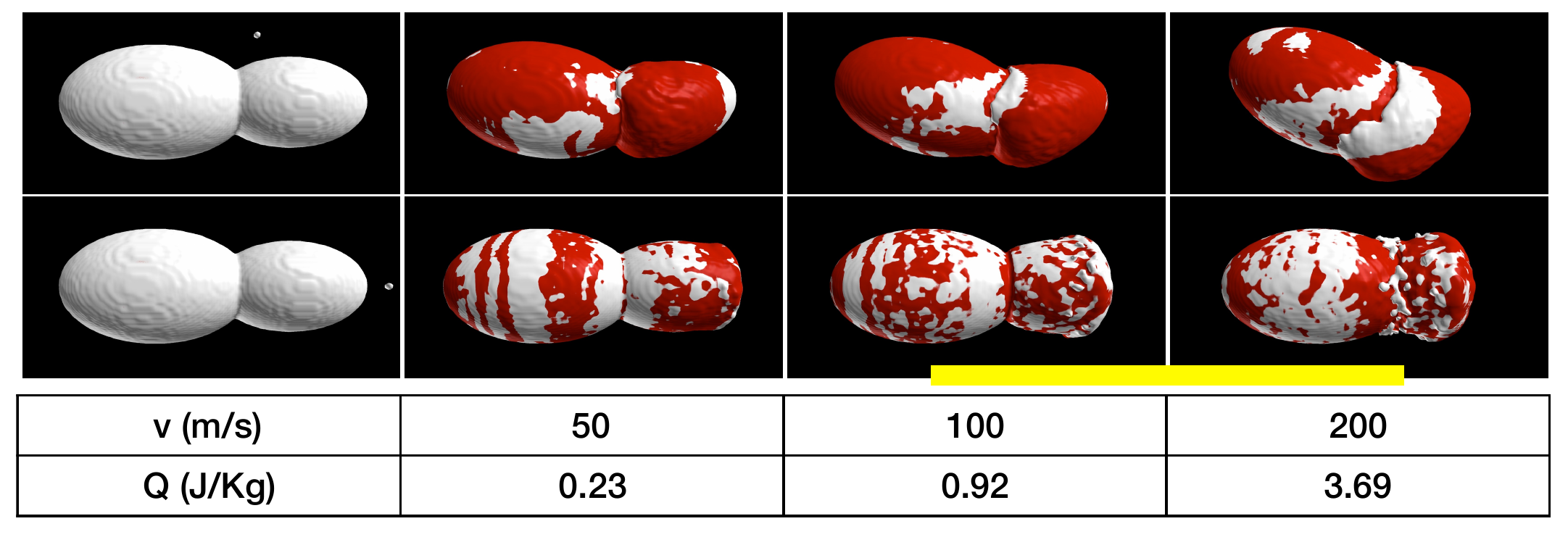}
   \caption{Results of impacts on generic bi-lobe shapes with nominal strength properties ($Y_T$ = 100 Pa) for two different impact geometries. $R_p$ = 100 m. The minimal specific energy required to cause a significant shape change is estimated as $Q_{bil} \sim$ 2.0$\pm$1.0 J/kg.}
   \label{fig:qbil}
\end{figure}

The results of our modeling of impacts on generic bi-lobe shapes (using nominal strength properties) are displayed in Figure  \ref{fig:qbil}.  The estimated minimal specific impact energies for reshaping are $Q_{bil} \sim$ 2.0$\pm$1.0 J/kg, which is slightly higher than in the case of the 67P/C-G-like shape with the same strength ($Q_{bil}$ [$Y_T$ = 100 Pa] $\sim$ corresponds to $Q_{reshape}$ for the  $Y_T$ = 1000 Pa case). 

\subsubsection{Catastrophic disruption threshold}\label{sec:qdresults}
\begin{figure}[h!]
   \centering
   \includegraphics[width=1.0\hsize]{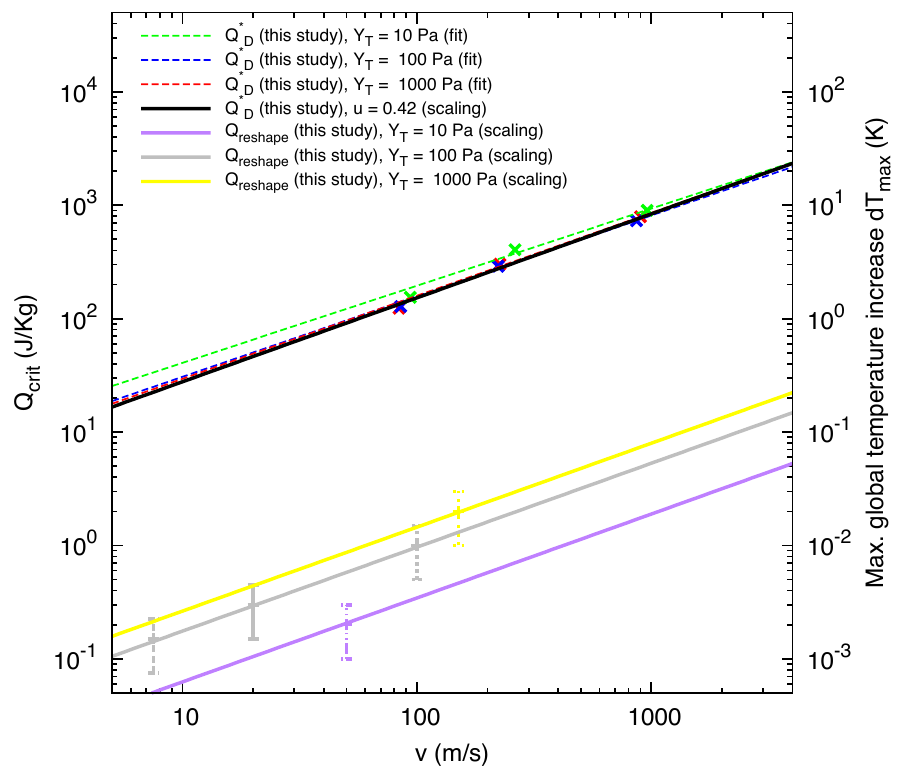}
   \caption{Critical specific impact energies $Q_{crit}$. The energy thresholds for shape-changing impacts on a 67P/C-G-like shape ($Q_{reshape}$) for different material strength are shown, as well as the catastrophic disruption energies $Q^*_D$ for various impact velocities. We note that the $Q_{bil}$ values found for shape-changing collisions on generic bi-lobe shapes overlap the results for  $Q_{reshape}$ with $Y_T$ = 1000 Pa; they are not shown seperately. The solid lines  show the scaling law (Equation \ref{eq:qscaling}) with parameters given in Table \ref{table:scaling}. 
   The maximal global temperature increase $dT$ shown on the right y-axis is estimated by assuming that all kinetic impact energy is converted into internal energy: $dT=Q_{crit}/c_p$ where a constant heat capacity $c_p$= 100 J/kg/K is used.}
   \label{fig:q_v_disr}
    \end{figure}

The results of our modeling of catastrophic disruptions of spherical bodies with the same mass and material properties as in the model of comet 67P/C-G are shown in Figure \ref{fig:q_v_disr}. We define the specific impact energy as
\begin{equation}
Q = 0.5 \mu_r V^2 / (M_t+M_p)
\end{equation}\label{eq:qdef}
where $\mu_r=M_p M_t /(M_t+M_p)$ is the reduced mass, $M_p$ is the mass of the projectile and $V$ the impact velocity. 
For the oblique (45$^\circ$) impacts considered here, we also take into account that only a part of the mass of the colliding bodies is interacting \citep{Leinhardt:2012ls}, and compute the $Q^*_D$ values of the equivalent head-on collisions.

As expected, the energy threshold for catastrophic disruption $Q^*_D$ $>>$ $Q_{reshape}$, by $\sim$ two orders of magnitude.

As found in previous studies (e.g. \citealp{Jutzi:2015gb}), in the disruption regime, the results for $Q^*_D$ are almost independent of the material (tensile) strength.

Our values of $Q^*_D$ for different impact velocities (Figure \ref{fig:q_v_disr}) agree well with scaling law predictions \citep{Housen:1990hh}, adopting a value for the coupling parameter of $\mu$ = 0.42, which is typical for porous materials.

   \begin{figure}[h!]
   \centering
   \includegraphics[width=1.0\hsize]{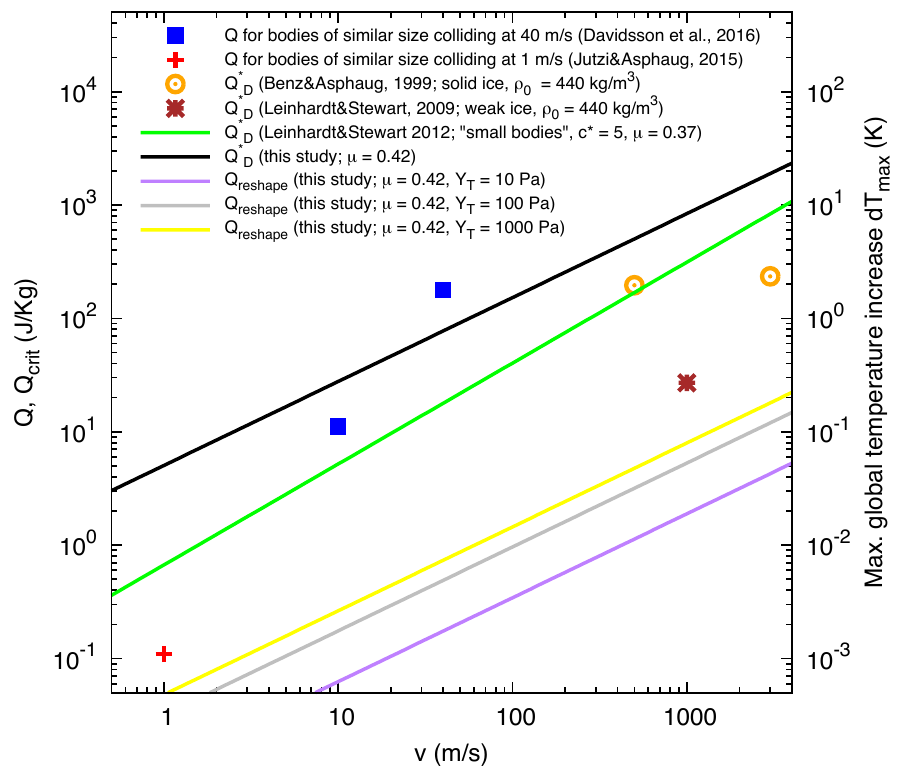}
   \caption{Comparison of  critical specific impact energies $Q_{crit}$. The scaling laws shown Figure \ref{fig:q_v_disr} are compared here with $Q^*_D$ values found in previous studies \citep{Benz:1999cj,Leinhardt:2009ic,Leinhardt:2012ls}. Also displayed are the specific energies $Q$ of collisions involving bodies of similar size (mass ratio of 1:2) for the bi-lobe forming collisions in study by \citet{Jutzi:2015ja} with very low velocities ($V\sim V_{esc} \sim$ 1 m/s) as well as for collisions with a velocity of $V$ = 40 m/s, corresponding to the average random velocity in the primordial disk during the first 25 Myr in the model by \citet{Davidsson:2016ds}.}
   \label{fig:q_v_disr_comp}
    \end{figure}

The $Q^*_D$ values for the weak, highly porous bodies considered here are slightly higher than the specific energies $Q^*_{D,ice}$ found for \emph{solid} bodies made of strong ice \citep{Benz:1999cj} (Figure \ref{fig:q_v_disr_comp}). This result reflects the dissipative properties of material porosity and is consistent with previous studies (e.g. \citealp{Jutzi:2010bf}).

Also shown in Figure \ref{fig:q_v_disr_comp} is the value of  $Q^*_D$ suggested  by \citet{Leinhardt:2009ic} for weak ice as well as $Q^*_D$ predicted from scaling laws for collisions between gravity-dominated bodies \citep{Leinhardt:2012ls}. In these studies, the effects of material porosity were not taken into account.

Finally, we also display in Figure  \ref{fig:q_v_disr_comp} the specific energies $Q$ involved in collisions of bodies of similar size (mass ratio 1:2) in the bi-lobe forming low-velocity regime investigated by \citet{Jutzi:2015ja}. As expected, those low-velocity ($V\sim V_{esc}$) accretionary collisions have specific impact energies far below the disruption threshold. For reference, we also show the specific energy for collisions with much higher velocities (v = 40 m/s), which correspond to the average random velocity in the initial primordial disk in the model by \citet{Davidsson:2016ds}. For a mass ratio of 1:2, the specific impact energies are even above energy threshold for catastrophic disruptions $Q^*_D$.

\subsection{Scaling laws for critical specific energies}
The results obtained in the previous section allow us to derive a $Q^*_D$ scaling law for porous cometary nuclei, which is a function of impact velocity $V$ and target size $R$ \citep{Housen:1990hh}: 
\begin{equation}\label{eq:qscaling}
Q^*_D= a R^{3\mu}V^{2-3\mu}
\end{equation}
where $\mu$ and $a$ are scaling parameters.

For $Q_{reshape}$ and $Q_{bil}$, we use a fixed target size $R$ = 1800 m. As shown in Figure \ref{fig:q_v_disr},  $\mu$ = 0.42  also reproduces well the velocity dependence of these critical specific energies.  
The scaling parameters for $Q^*_D$, $Q_{reshape}$  and $Q_{bil}$ are given in Table \ref{table:scaling}. 

\begin{table}
\caption{Parameters (SI units) for the scaling law $Q_{crit} = a R^{3\mu}V^{2-3\mu}$, where $R$ is the target radius and $V$ the impact velocity. The scaling for shape-changing impacts on 67P/C-G ($Q_{reshape}$) and for impacts on generic bi-lobe shapes ($Q_{bil}$)  only hold for a fixed size ($R$ = 1800 m).}            
\label{table:scaling}      
\centering                        
\begin{tabular}{l c c }        
\hline\hline              
Scaling & $\mu$ & $a$ \\    
\hline                        
   $Q^*_D$ & 0.42 & 4.0e-4 \\     
   $Q_{reshape}$ (10 Pa) & 0.42 & 9.0e-7 \\  
   $Q_{reshape}$ (100 Pa; nominal) & 0.42 & 2.5e-6 \\
    $Q_{reshape}$ (1000 Pa) & 0.42 & 3.8e-6 \\
    $Q_{bil}$ (nominal) & 0.42 & 3.8e-6 \\
   \hline                               
\end{tabular}
\end{table}

\begin{figure}[]
\centering
\includegraphics[width=1.0\hsize]{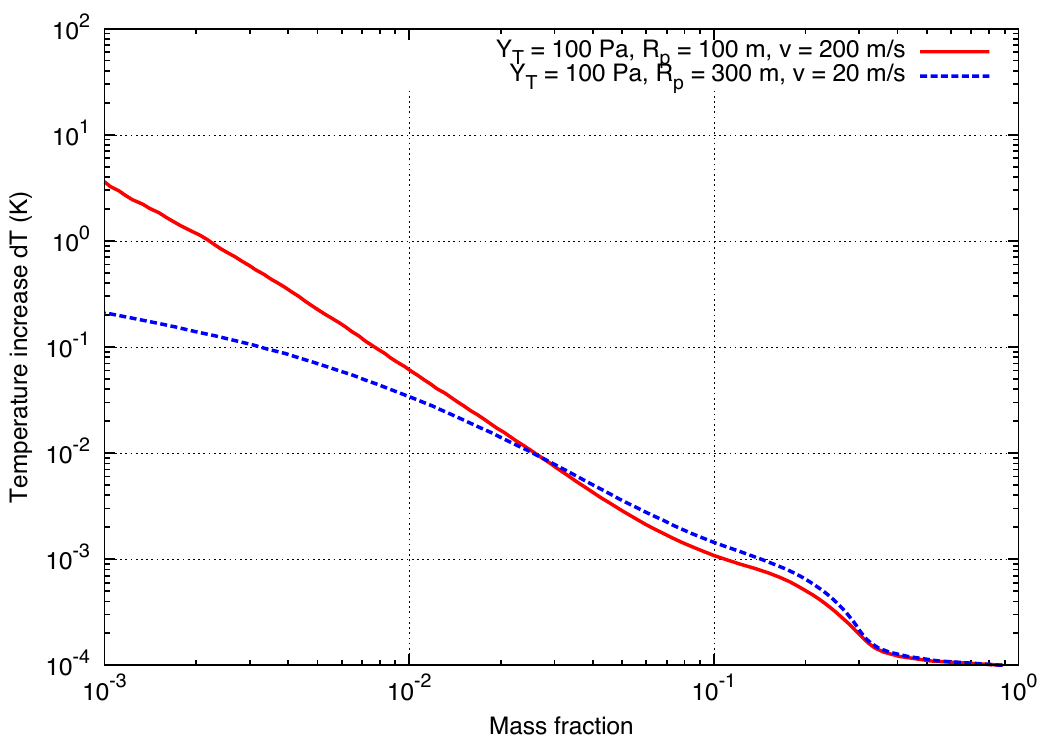}
\caption{Cumulative post-impact temperature increase $dT$  for two specific cases of shape-changing collisions, as indicated in the plot.}
\label{fig:m_dT}
\end{figure}
    
 \begin{figure}[h!]
\centering
\includegraphics[width=\hsize]{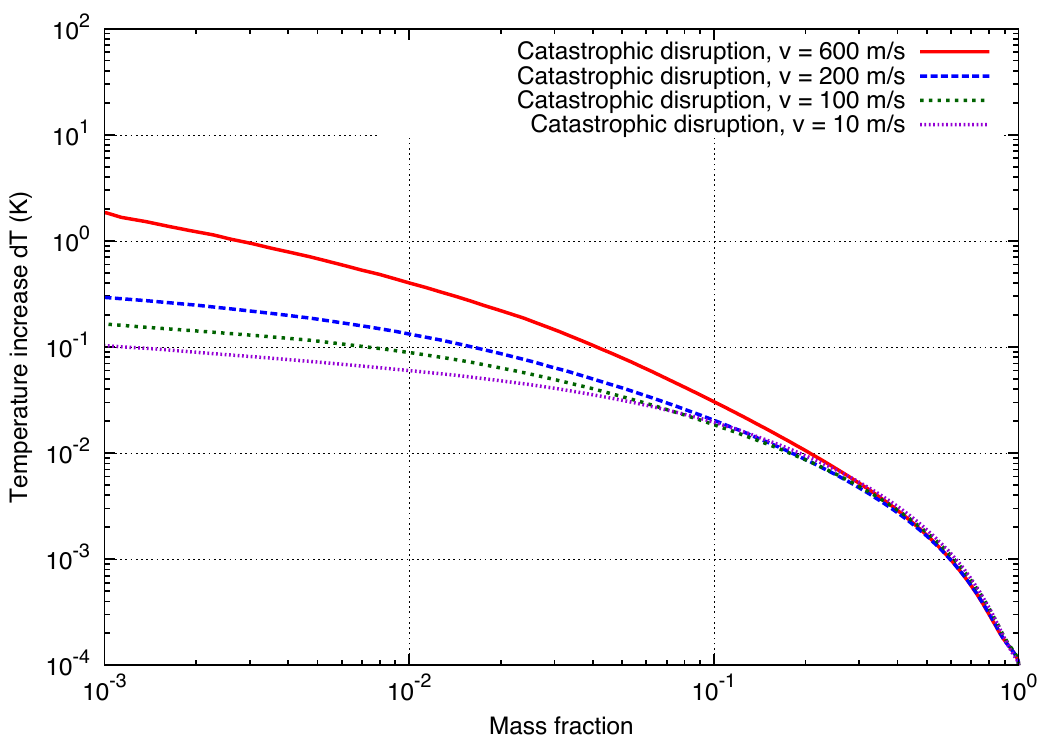}
\caption{Fraction of material in the final body that experienced a temperature increase larger than a certain value $dT$ in catastrophic disruptions with different velocities $V$. The mass of the largest remnant $M_{lr}/M_{t} \sim 50\%$. Only the material belonging (i.e. which is bound) to the largest remnant is considered in the analysis.}
 \label{fig:m_dT_disr}
 \end{figure}    
 
\subsection{Impact heating}
The effects of the impacts considered in this study  (shape-changing impacts as well as catastrophic disruptions) are analyzed in terms of impact heating and porosity evolution (below). First, in order to get an idea of the maximal global heating, we simply convert the total specific impact energy into a global temperature increase $dT=Q_{crit}/c_p$  where a constant heat capacity $c_p$ = 100 J/kg/K is used. The value of $c_p$ is a rough mass weighted average of the heat capacity of ice \citep{Klinger:1981} and silicates \citep{Robie:1982rh} at low temperatures ($\sim$ 30 K), assuming a dust-to-ice mass ratio of 4 \citep{Rotundi:2015rs}. Figures \ref{fig:q_v_disr} and \ref{fig:q_v_disr_comp} (scale on the right) shows these $dT$ values corresponding to collisions with a given specific impact energy. 

From this simple estimation, it is already obvious that impacts with energies comparable to $Q_{reshape}$, the maximal global temperature increase must be limited to small values ($dT << 1$ K). On the other hand, catastrophic impacts at kilometer scales may have the potential to lead to significant large scale heating, depending on how the impact energy is distributed.  

We compute the actual post-impact  $dT$ distribution for a few specific cases of the shape-changing (section \ref{sec:shaperesults}) as well as catastrophic collisions (section \ref{sec:qdresults}). In the later, we only consider the material which ends up in the largest remnant ($\sim 50\%$ of the initial target mass). The cumulative temperature distribution in the case of the shape-changing impacts (Figure  \ref{fig:m_dT}) confirms that only a very small fraction of the material experiences significant heating. For the catastrophic collisions we find that the part of the target which experiences the largest impact effects is ejected. As a result, the material which is bound to the largest remnant (consisting $\sim 50\%$ of the original target mass) is not affected much be the collision (Figure  \ref{fig:m_dT_disr}) and the heating is limited ($< 1\%$ of the mass is heated by $dT >$ 1 K), even at relatively high collision velocities (600 m/s).

\subsection{Porosity evolution}
Porosity is changed by impacts in multiple ways. First, material is compacted due to the pressure wave generated by the impact. On the other hand, material is ejected and the process of reaccumulation of the gravitationally bound material can give rise to additional macroporosity. Our porosity model computes the degree of compaction (change of the distention variable). In order to specify the increase of the macroporosity, we treat each SPH particle individually according to its ejection/reaccumulation history. Particles which are lifted off the surface or are  ejected and reaccumulated experience a density decrease, resulting in an increase of porosity. We assume that  reaccumulated material can lead to the addition of macroporosity of maximal 40\%, a typical porosity of rubble-pile asteroids \citep{Fujiwara:2006fk}. To compute the total final porosity $\phi_{total}$ resulting from compaction and reaccumulation, we use the relation
\begin{equation}
\phi_{total} = 1-1/\alpha_{total}
\end{equation}
and define the distention
\begin{equation}
\alpha_{total} = \mbox{min}(\rho_{compact}/\rho_{min},\alpha_{max})
\end{equation}
where $\rho_{min}$ is the minimal density reached by the SPH particle and $\rho_{compact}$ = 1980 kg/m$^3$. For this calculation we consider all particles which are gravitationally bound to the main body (largest remnant). The upper limit of the distention is given by
\begin{equation}
\alpha_{max}=\alpha_{0}\alpha_{v}
\end{equation}
where $\alpha_{v}$ is the distention value corresponding to 40\% macroporosity, $\alpha_{v}=(1-\phi_v)^{-1}$ with $\phi_v=0.4$, and $\alpha_0$ = 4.5 is the initial distention.

\begin{figure}
   \centering
   \includegraphics[width=1.0\hsize]{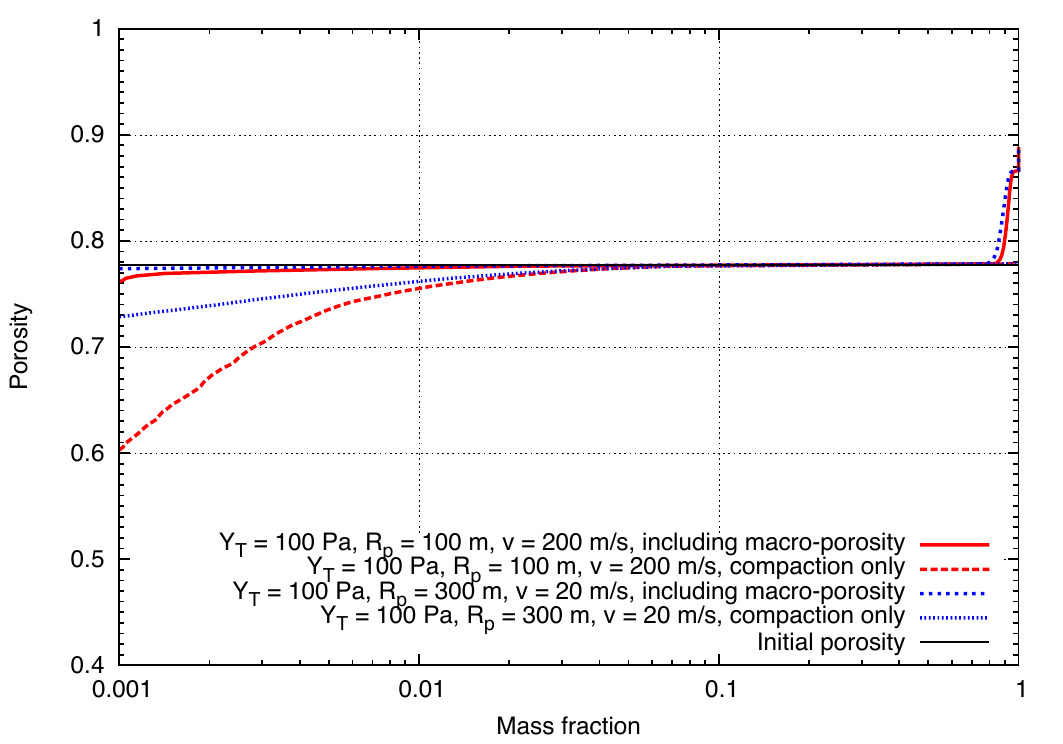}
   \caption{Post-impact porosity distribution for two specific cases of shape-changing collisions, as indicated in the plot. The porosity calculation takes into account compaction as well as the increase of macroporosity.  For comparison, the porosity distributions resulting from compaction only are shown as well. The final average porosity (compaction plus addition of macroporosity by reaccumulation) is 78.8\% ($R_p$ = 100 m) and 79.2\% ($R_p$ = 300 m), respectively, while the initial porosity was 77.8\%.}
   \label{fig:m_dist}
    \end{figure}

\begin{figure}
\centering
\includegraphics[width=\hsize]{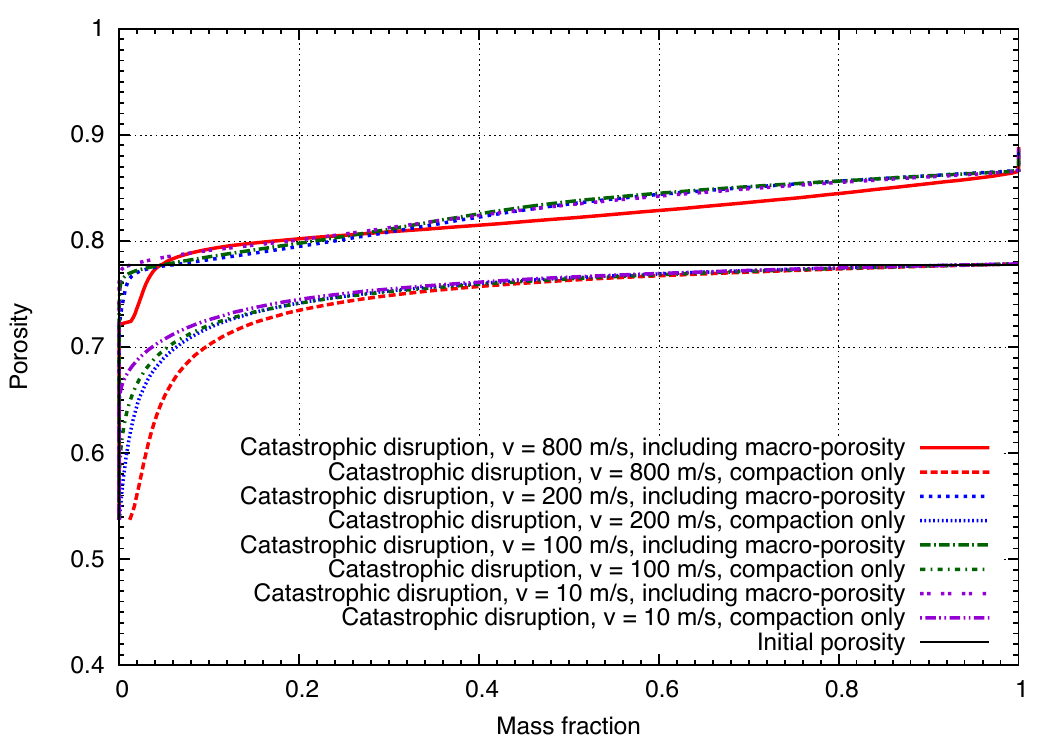}
\caption{Same as Figure \ref{fig:m_dist} but for catastrophic collisions. $M_{lr}/M{tot} \sim 50\%$; only the material bound to the largest fragment $M_{lr}$ is considered. The final average porosity (compaction plus addition of macroporosity by reaccumulation) is 83.3\% ($V$ = 10 m/s), 83.3\% ($V$ = 100 m/s), 83.3\% ($V$ = 200 m/s), 82.4\% ($V$ = 800 m/s), respectively, while the initial porosity was 77.8\%.}
 \label{fig:m_dist_disr}
 \end{figure}

The resulting cumulative porosity distributions are calculated for the same cases of shape-changing and catastrophic collisions as discussed in the previous section (Figures   \ref{fig:m_dist} and \ref{fig:m_dist_disr}). In the case of the shape-changing collisions, compaction is quite limited, even though the impacted lobe is severely disrupted. Because of the very low gravity, material is lifted off the surface/ejected by the impact. Due to the addition of macroporosity resulting from reaccumulation, the final average porosity is about the same as the initial porosity (Figure \ref{fig:m_dist}).

In the catastrophic disruptions, most of the material which undergoes collisional induced compaction does not remain on the main body (largest remnant). As a result, only $\sim$ 10\% of the material in the final main body has experienced significant compaction. On the other hand reaccumulation plays a major role in this collision regime, resulting in a significant increase of macroporosity. The final porosity is therefore even slightly higher than the initial porosity (Figure \ref{fig:m_dist_disr}).

In Paper II, the interior porosity distribution of bi-lobe structures resulting from sub-catastrophic collisions are compared to observations of comet 67P/C-G.

\section{The combined dynamical and collisional evolution of comet 67P/C-G}\label{sec:dynmodel}

\subsection{Modeling approach}

We follow the approach described in \citet{Morbidelli:2015vm} in order to combine the dynamical evolution of the planetesimals precursors of Jupiter family comets with their collisional evolution. We do not repeat here a detailed description of the procedure, but we report on the differences and the improvements in the new implementation. 

These are of three kinds. First, we consider here only the dynamical dispersal of the original trans-Neptunian disk of planetesimals, which generates the Scattered Disk (the current source reservoir of JFCs). Thus, we neglect the phase ranging from the time when the gas was removed from the protoplanetary disk to the time when the giant planets developed a dynamical instability that dispersed the planetesimal disk \citep{Tsiganis:2005tg,Gomes:2005gl}. This choice is made because \citet{Morbidelli:2015vm} already showed that in the standard model, a comet the size of 67P/C-G has no chance to survive intact during this phase, if protracted for $\sim 400$ My. On the other hand the debate on the timing of the giant planet instability is still open \citep[see for instance][]{Kaib:2016kc,Toliou:2016tm}, so it might be possible that the aforementioned phase is short. There is no doubt, however, that the dispersal of the trans-Neptunian disk occurred and that this formed the Scattered Disk. In this case, \citet{Morbidelli:2015vm} showed that during this process the number of catastrophic collisions for planetesimals the size of 67P/C-G is $\sim 1$, so there might be some objects escaping break-up events. Thus, in this work we focus on this case, using improved assessments on the specific energies for catastrophic break-up and for reshaping, described in the previous sections. 

The second improvement over  \citet{Morbidelli:2015vm} concerns the dynamical simulations.  \citet{Morbidelli:2015vm} used the simulation of Gomes et al. (2005), which covered only the first 350 My after the giant planet instability. This is when most of the action happens, but the subsequent 3.5-4.0 Gy cannot be neglected.  \citet{Morbidelli:2015vm} assumed that over this remaining time the orbital distribution of the Scattered Disk does not evolve any more, but its population decays exponentially down to 1\% of the original population after 4 Gy. The 1\% fraction comes from previous studies of the long term evolution of the Scattered Disk \citep{Duncan:1997dl}. Here we use the simulations presented in \citet{Brasser:2013dw}, which constitute a much more coherent set.  \citet{Brasser:2013dw} studied the dispersal of the trans-Neptunian planetesimal disk during the giant planet instability using a large number of particles (1,080,000; including clones). At the end of the instability, they drove the giant planets towards their exact current orbits, so to avoid artefacts in the subsequent long-term evolution of the Scattered Disk. The evolution of the Scattered Disk was followed for 4 Gy. Because the number of active particles decays over time, the test particles have been cloned 3 times, at 0.2, 1.0 and 3.5 Gy. In the final 0.5 Gy simulation, the particles leaving the Scattered Disk to penetrate into the inner solar system as JFCs have been followed, in order to compare their orbital distribution with that of the observed comets. This final step is crucial to demonstrate that the Scattered Disk generated from the dispersal of the trans-Neptunian disk by the giant planet instability is a valid source of JFCs. 

The third improvement over  \citet{Morbidelli:2015vm} is that the collisional evolution is followed only for the particles that eventually become JFCs in the final 0.5 Gy simulation. These are 87 particles. We think that, potentially, this is an important improvement. The particles that penetrate the inner solar system at the present time might have had specific orbital histories relative to the other particles that either became JFCs much earlier or are still in the Scattered Disk today. Averaging the collisional histories of these three categories of particles may not be significant to address the specific case of 67P/C-G, which clearly became JFC only in recent time.

Like in  \citet{Morbidelli:2015vm} the number of collisions suffered by each considered body is computed assuming that the initial disk particles represent a population of $2\times 10^{11}$ planetesimals with diameter $D>2.3$ km, with a differential size distribution characterized by an exponent $q$. The minimum projectile size is determined by the scaling law (equation \ref{eq:qscaling}) for the critical specific energy, with parameters given in Table \ref{table:scaling}. 
As for the exponent $q$, in agreement with  \citet{Morbidelli:2015vm} and previous studies of the comet size distribution, we consider here the cases with $q=-2.5, -3.0$ and $-3.5$. However, in the meantime the New Horizons mission to Pluto and Charon has measured the crater size frequency distribution, allowing the assessment of the size distribution of the trans-Neptunian objects larger than a few km in diameter \citep{Singer:2015ss}. The preliminary results\footnote{We note that based on the most recent results it has been suggested that there may be a deficit of small objects \citep{Singer:2016sm}; see discussion in section \ref{sec:validity}.} suggest $q=-3.3$. Thus, we consider the results for $q=-3.0$ and $-3.5$ as the most significant. However, we note that in alternative models  \citep{Davidsson:2016ds} shallower slopes are preferred. 

We stress that the approach followed in  \citet{Morbidelli:2015vm} and in this work is conservative. This means that the number of collisions that are estimated is a lower bound of the real number of collisions. This is because the number of bodies assumed in the initial trans-Neptunian disk ($2\times 10^{11}$ with $D>2.3$ km) is the minimum required, in absence of collisional comminution, to form an Oort cloud and a Scattered disk that contain enough objects to be sufficient sources of the LPC and JFC fluxes that we observe today.  

 \begin{figure}
 \centering
 \includegraphics[width=1.0\hsize]{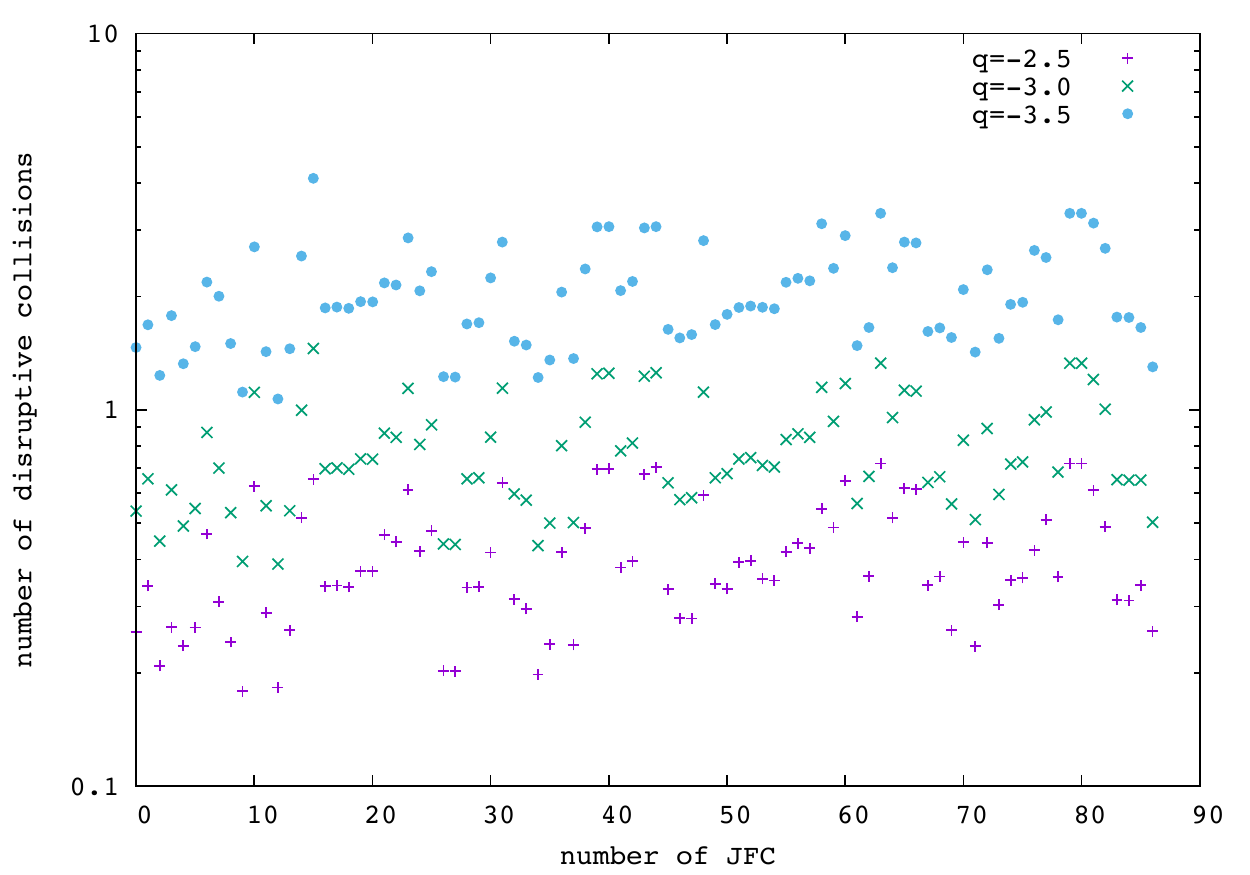}
 \caption{The number of expected catastrophic collisions $N_{disrupt}$ during the formation and evolution of the Scattered Disk for the particles that eventually become JFCs in the final 0.5 Gy simulation. $N_{disrupt}$ is computed using the scaling parameters for our new $Q^*_D$. The symbols depict different values for the exponent of the differential size distribution $q$, as labeled in the plot.}
 \label{fig:nb_col_qd}
\end{figure}

 \begin{figure}[h!]
 \centering
 \includegraphics[width=1.0\hsize]{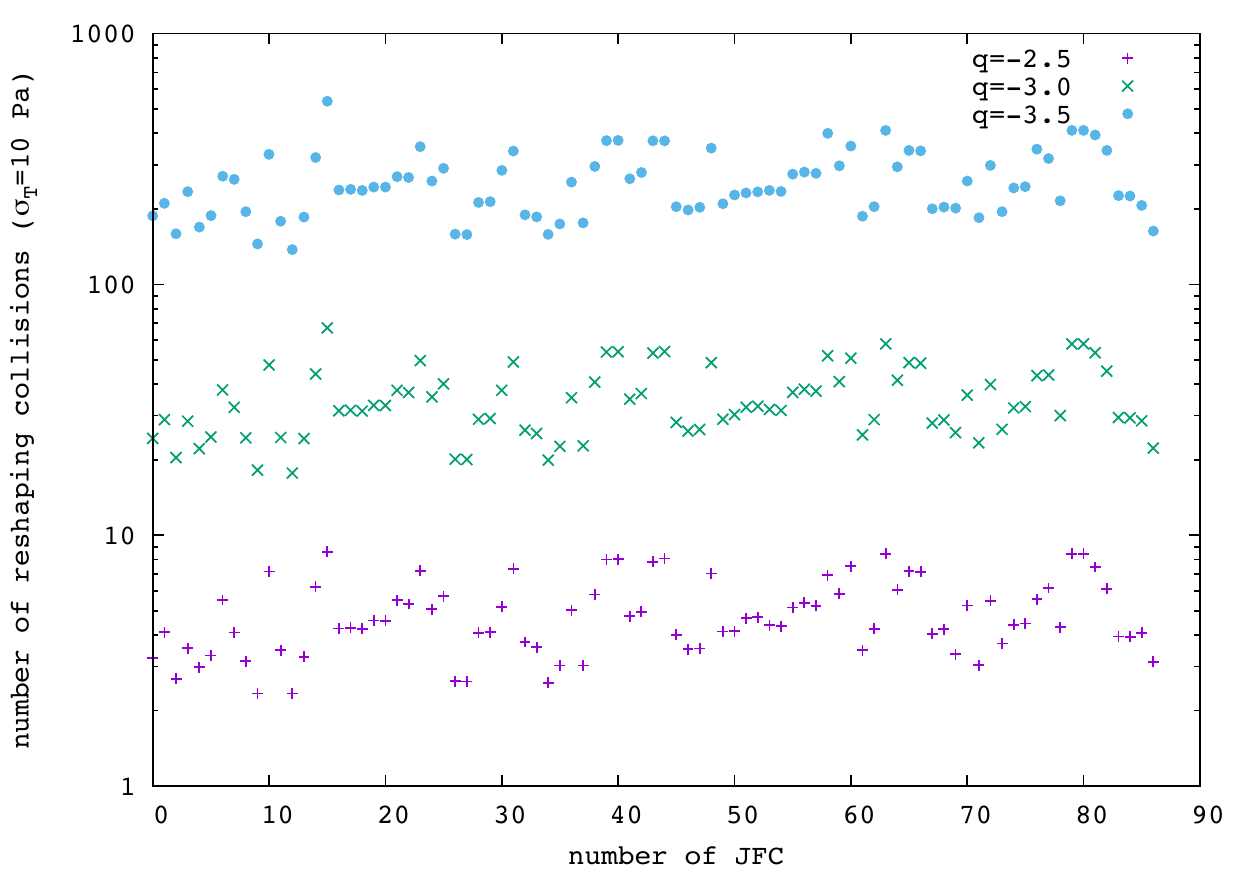}
  \includegraphics[width=1.0\hsize]{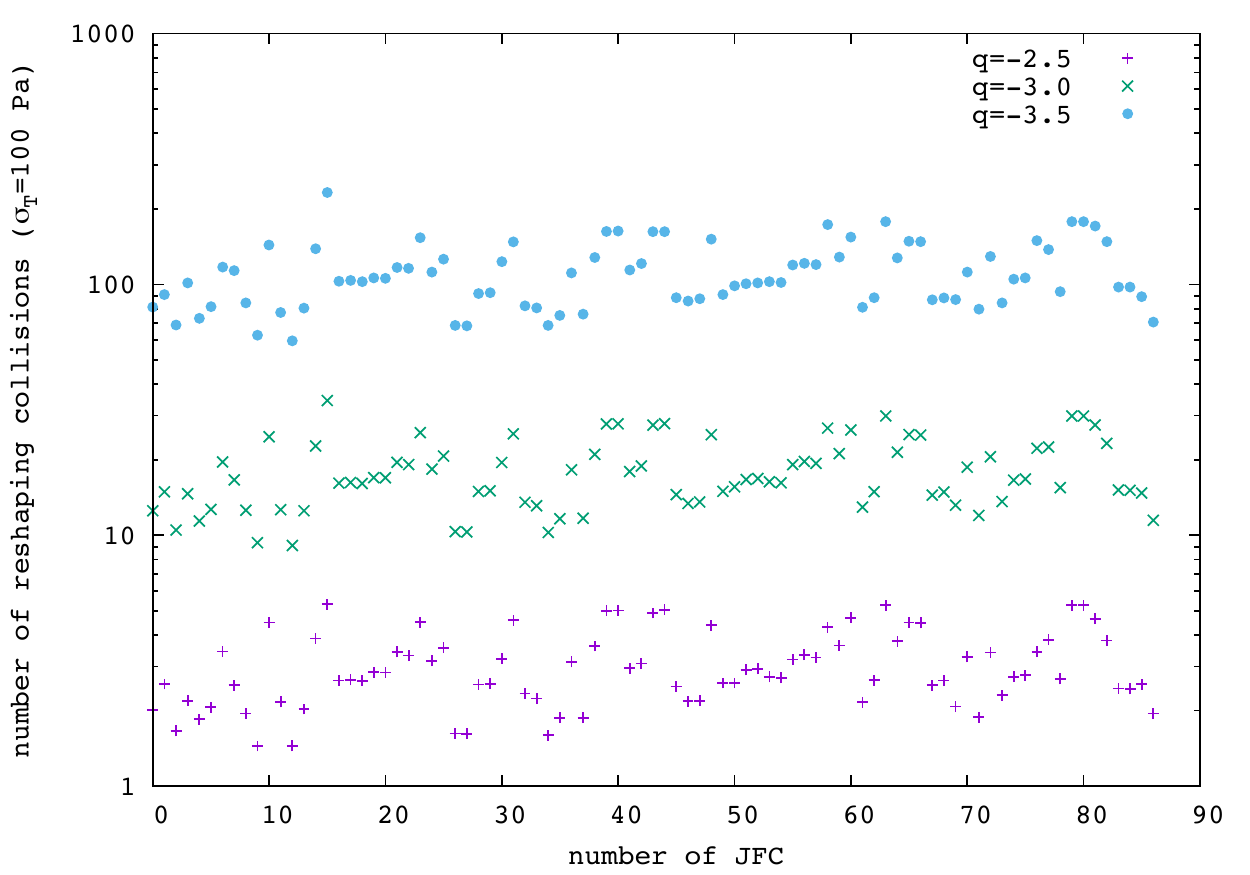}
   \includegraphics[width=1.0\hsize]{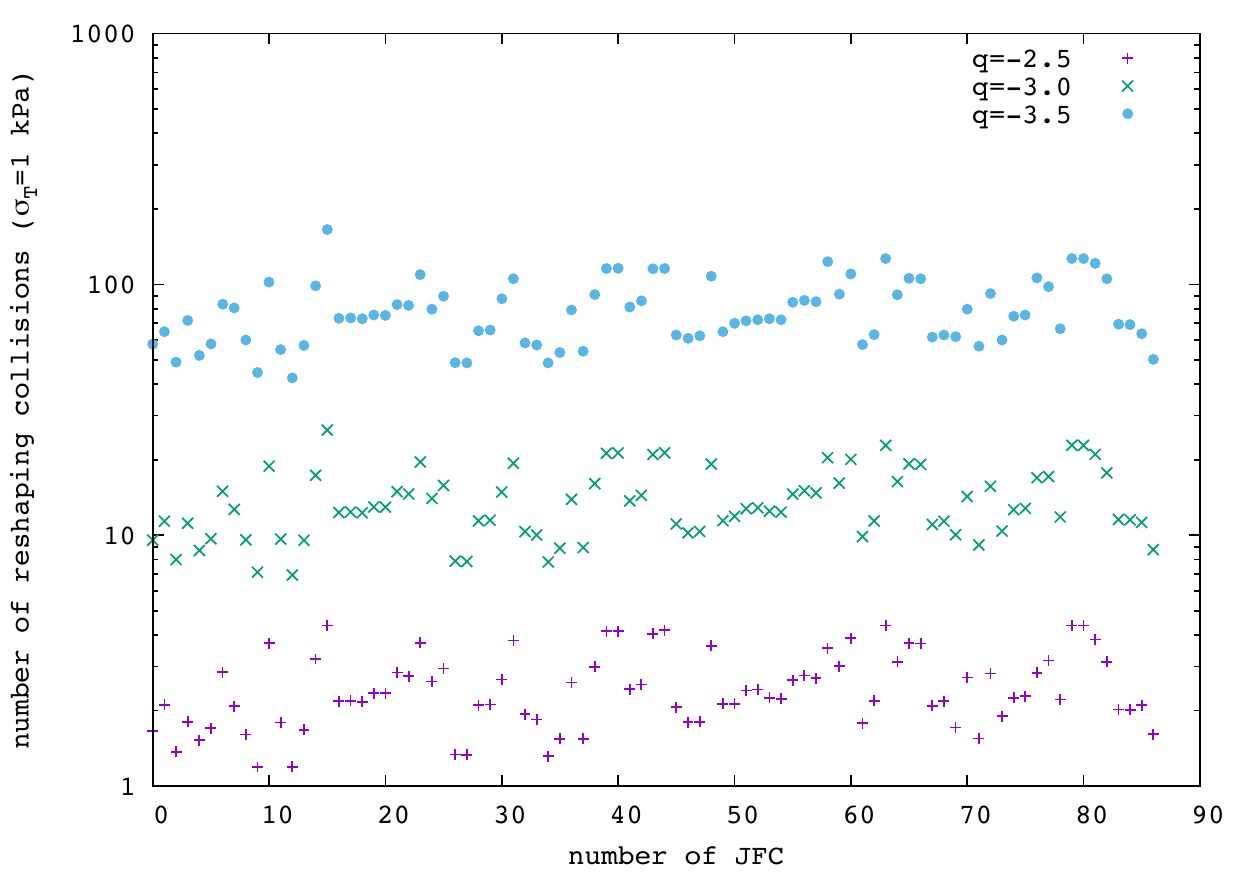}
 \caption{Same as Figure \ref{fig:nb_col_qd}, but shown is the number  of shape-changing collisions on a 67P/C-G-like body $N_{reshape}$, computed using scaling the parameters for $Q_{reshape}$ for different strengths. We note that the number of shape-changing collisions $N_{bil}$ in the case of a generic bi-lobe shape with nominal strength properties is the same as $N_{reshape}$ for $Y_T$ = 1000 Pa (shown in the plot at the bottom).}
 \label{fig:nb_col_shape}
\end{figure}

\begin{figure}[h!]
 \centering
 \includegraphics[width=1.0\hsize]{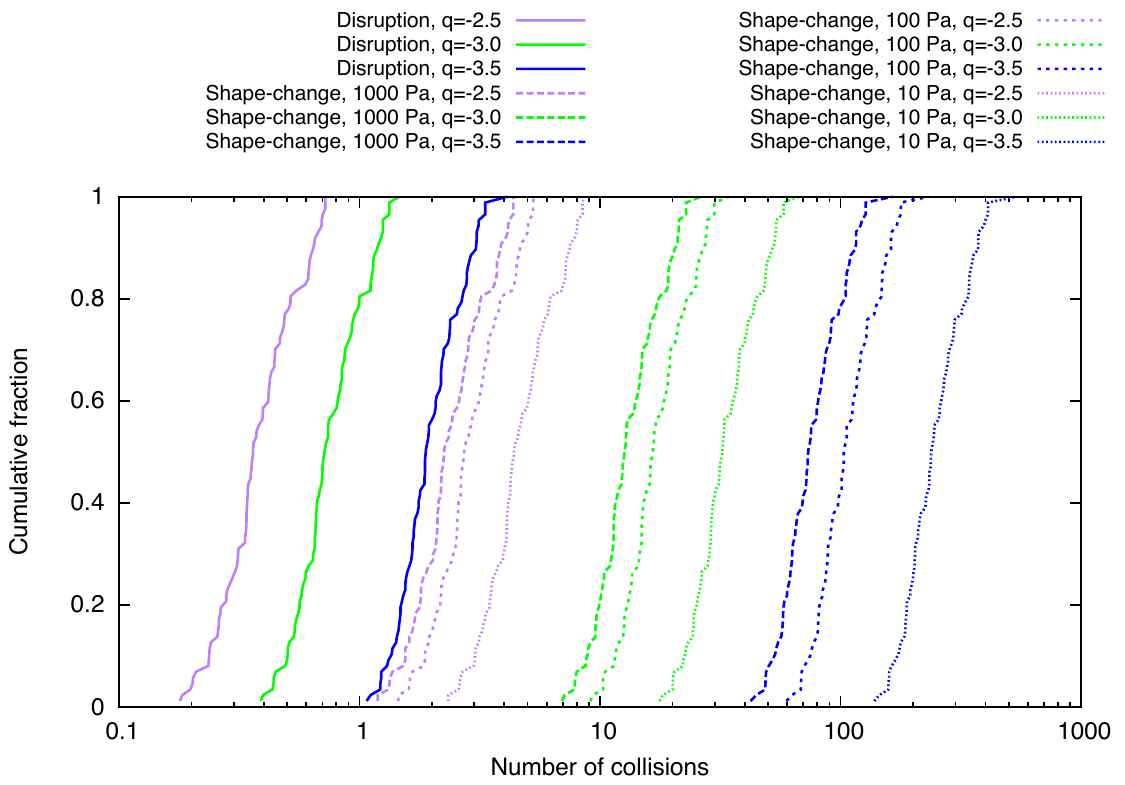}
 \caption{Cumulative fraction of particles (that eventually become JFCs) as a function of the number of collisions. This is an alternative representation of the results already shown in Figs. \ref{fig:nb_col_qd} and  \ref{fig:nb_col_shape}. The solid lines correspond to the  $Q^*_D$ scaling; the dotted lines were computed using $Q_{reshape}$ (for different strength values).}
 \label{fig:nb_col_qd_nb_cumu}
\end{figure}

\begin{figure}[h!]
 \centering
 \includegraphics[width=1.0\hsize]{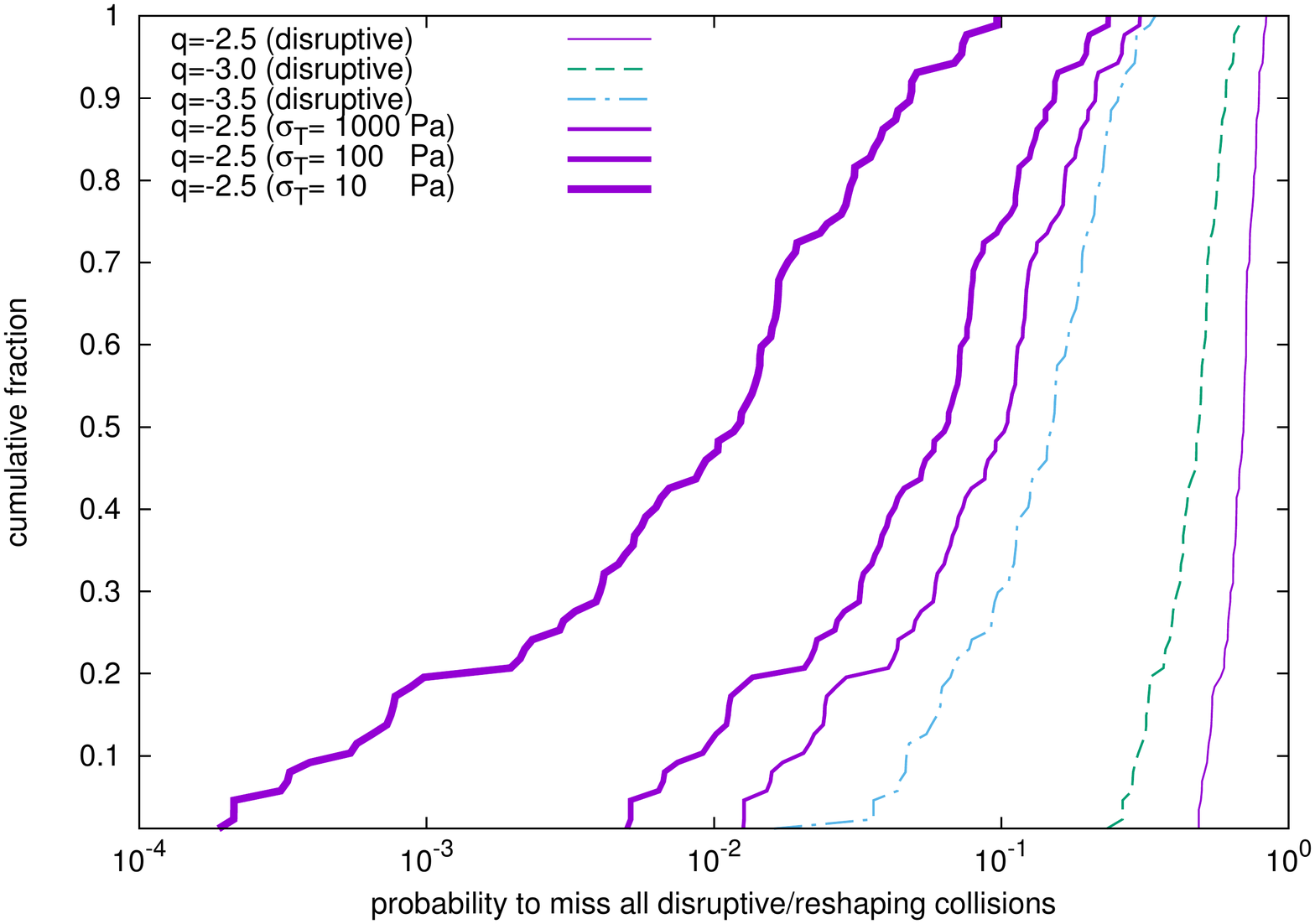}
 \caption{Cumulative fraction of particles (that eventually become JFCs) as a function of the probability $P(0)$ to escape all collisions. The different line styles refer to different exponents for the differential size distribution $q$, as labeled on the plot. The three curves on the right correspond to the  $Q^*_D$ scaling; the three curves on the left correspond to  $Q_{reshape}$ (with different strength values 10 Pa, 100 Pa and 1 kPa from left to right). For $q = -3.0$ and $q= -3.5$, the probability to miss all reshaping collisions is $P(0)<< 10^{-4}$ and the corresponding curves are not displayed here.}
 \label{fig:nb_col_qd_p_cumu}
\end{figure}

 \begin{figure}[h!]
 \centering
 \includegraphics[width=1.0\hsize]{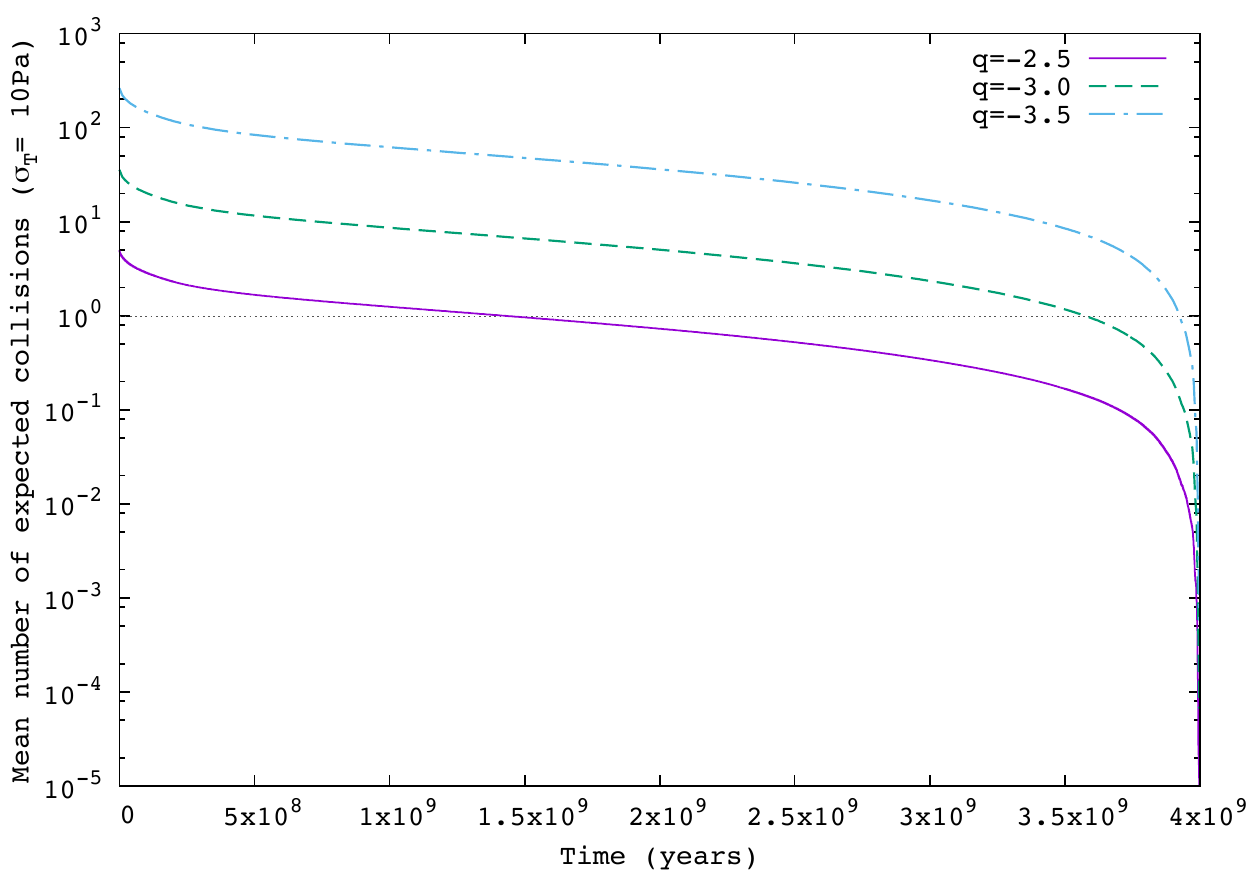}
  \includegraphics[width=1.0\hsize]{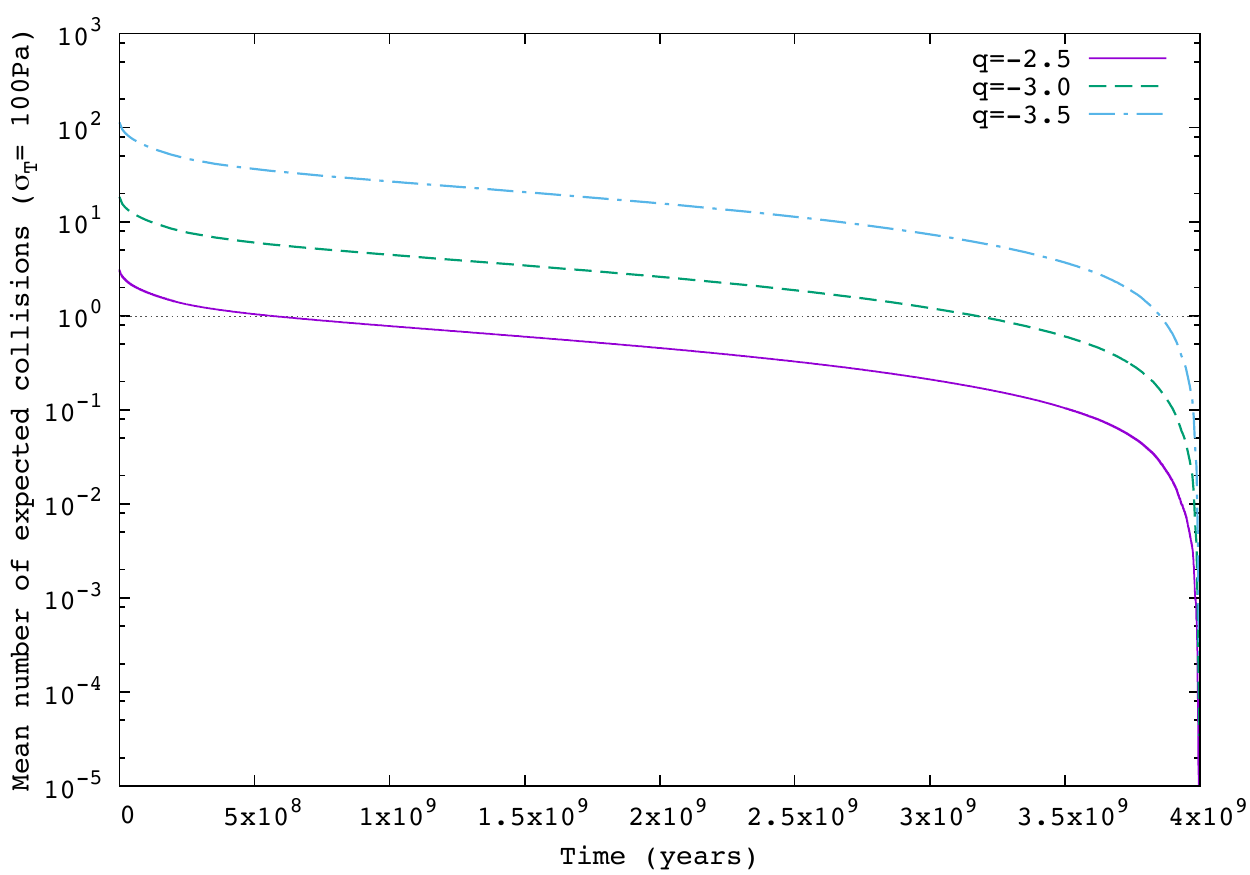}
   \includegraphics[width=1.0\hsize]{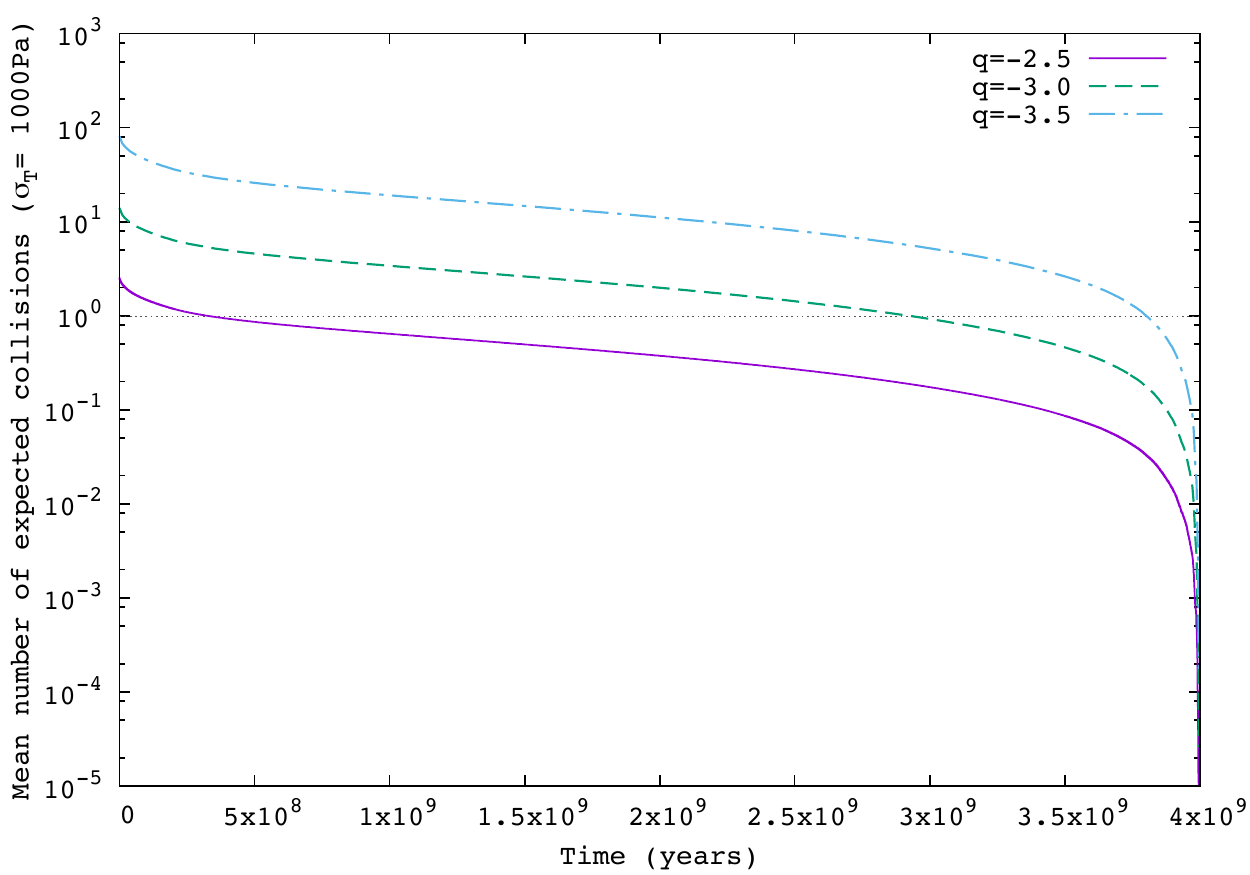}
 \caption{Mean number of reshaping collisions $N_{reshape}$ expected for 67P/C-G-like objects as a function of time for different strengths, as indicated on the $y$-axis. We note that the number of shape-changing collisions $N_{bil}$ in the case of a generic bi-lobe shape with nominal strength properties is the same as $N_{reshape}$ for $Y_T$ = 1000 Pa (bottom). Time $t = 0$ corresponds to the beginning of the dynamical dispersal of the original trans-Neptunian disk of planetesimals, which generates the Scattered Disk; $t = 4\times 10^9$ years is now.}
  \label{fig:nb_col_shape_time}
\end{figure}

 \begin{figure}[h!]
 \centering
 \includegraphics[width=1.0\hsize]{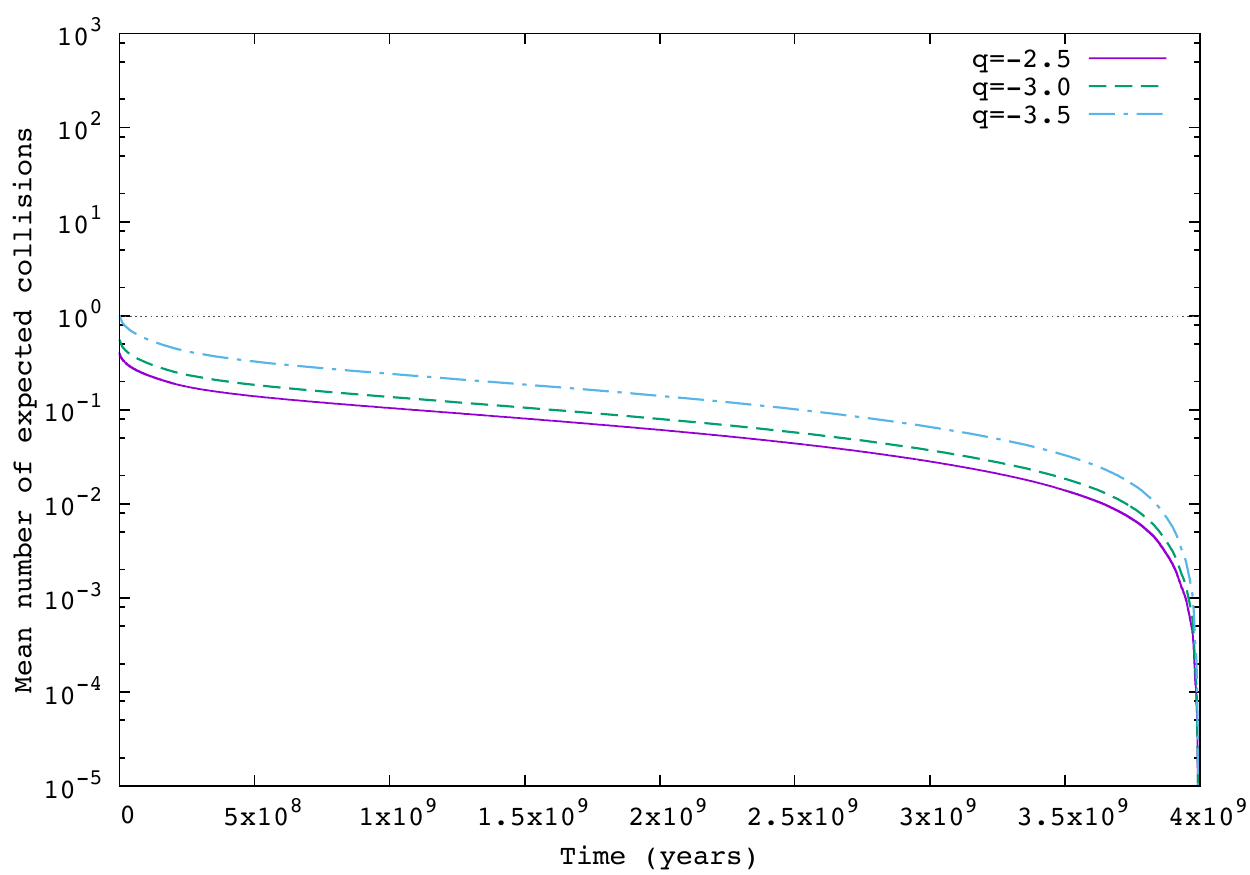}
  \includegraphics[width=1.0\hsize]{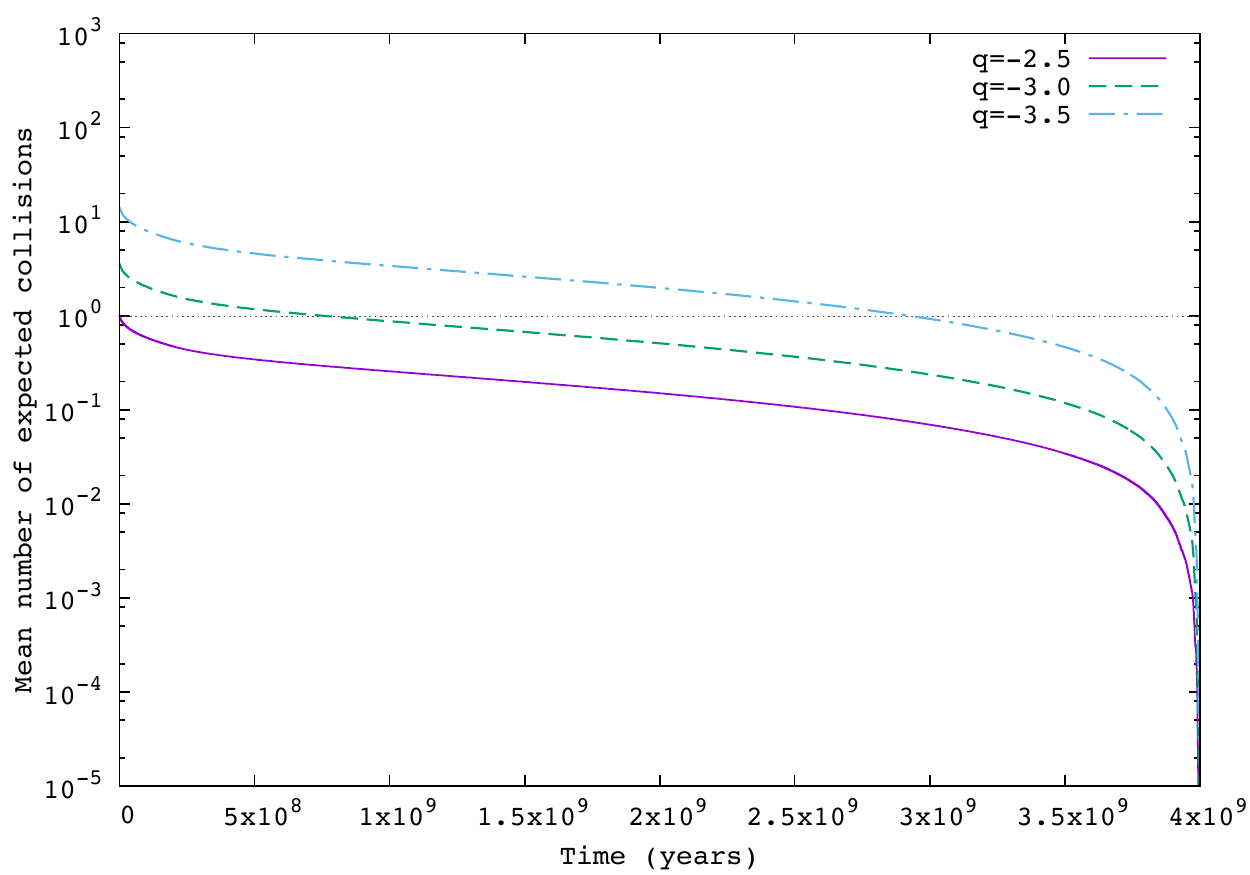}
  \caption{Top: Mean number of potential 67P/C-G-forming catastrophic collisions of a parent body of $R$ = 3km (computed using $Q^*_D$) as a function of time $t$ (defined as in Figure  \ref{fig:nb_col_shape_time}). Bottom: Same, but for the scenario of 67P/C-G formation by low energy sub-catastrophic collisions.}
  \label{fig:nb_col_form_time}
\end{figure}

\subsection{Results: number of disruptive and shape-changing collisions}

The number of events for each particle surviving in the Scattered Disk at the end of the disk dispersal simulation is shown in Figs.  \ref{fig:nb_col_qd} - \ref{fig:nb_col_shape}. The results for the various types of collisions, using the corresponding scaling laws ($Q^*_D$ and $Q_{reshape}$), are plotted. We note that the results for $Q_{bil}$ (impacts on generic bi-lobe shape using nominal material properties) are the same as in the case of $Q_{reshape}$ with $Y_T$ = 1000 Pa (Table \ref{table:scaling}); they are therefore not displayed separately. 

Compared to the results by  \citet{Morbidelli:2015vm}, the number of disruptive collisions  is smaller (Figure \ref{fig:nb_col_qd}). This is mainly due to the fact the new $Q^*_D$ scaling law used here leads to disruption energies which are higher than the ones by \citet{Benz:1999cj} (which were used in the previous study). As discussed in section \ref{sec:qdresults}, this can be explained by the highly dissipative properties of porous materials, which are taken into account in the new $Q^*_D$.  As Figure \ref{fig:nb_col_qd} shows, for shallow size distributions, it is possible in principle that a significant fraction of the 67P/C-G sized objects escaped all \emph{catastrophic} collisions. 

On the other hand, the number of \emph{shape-changing} collisions (Figure \ref{fig:nb_col_shape}), requiring a much smaller impact energy ($Q_{reshape}$), is substantially larger than the number of catastrophic events. As expected, the weaker the strength the larger the number of reshaping collisions taking place. Also, the steeper the size distribution (larger $q$), the larger the number of collisions happening. However, in any case, even for the largest strength (1000 Pa) and the shallowest slope ($q$ = -2.5), the number of reshaping collisions largely exceeds 1 for all comets. 

The results are summarized in Figure \ref{fig:nb_col_qd_nb_cumu} which shows the cumulative fraction of particles as a function of the number of collisions. In Figure \ref{fig:nb_col_qd_p_cumu},  the number of collisions $N_{coll}$ is converted into a probability to avoid all collisions $P(0) = \text{exp}(-N_{coll})$ and the normalized cumulative distribution of the $P(0)$ values is plotted. The average number of collisions and the related probabilities are given in Table \ref{table:ncoll}.

\begin{table*}
\caption{Average number of shape-changing collision on a 67P/C-G-like object  ($N_{reshape}$),  shape-changing collisions  on a generic bi-lobe body ($N_{bil}$) and catastrophic  collisions ($N_{disrupt}$). The corresponding probability $P(0)$ to avoid all collisions is given in parenthesis.}      
\label{table:ncoll}
\centering                        
\begin{tabular}{ | l  | c c  | c c | c c |}        
\hline  \hline              
Type & $q$ = -2.5 & $ $ & $q$ = -3 & $ $ & $q$ = -3.5 & \\  
\hline 
$N_{disrupt}$ &0.41 &(6.7E-1) &0.79 &(4.5E-1) &2.06 &(1.3E-1)\\
\hdashline 
\hdashline 
$N_{reshape}$ (10 Pa)&4.92 &(7.3E-3) &35.1 &(6.0E-16)&258 &(7.3E-113)\\
$N_{reshape}$ (100 Pa)&3.06 &(4.7E-2) &18.1 &(1.4E-8)& 112 &(2.4E-49)\\
$N_{reshape}$ (1000 Pa) & 2.53& (8.0E-2)&13.8 &(1.0E-6) &79.6& (2.7E-35)\\
\hdashline 
\hdashline 
$N_{bil}$ (nominal) & 2.53& (8.0E-2)&13.8 &(1.0E-6) &79.6& (2.7E-35)\\
   \hline                               
\end{tabular}
\end{table*}

It is also interesting to look at the number of collisions as a function of time (Figs. \ref{fig:nb_col_shape_time} and \ref{fig:nb_col_form_time})  as this in principle allows us to determine the time at which on average the last event of a certain type took place.  

For size distributions with $q\le -3.0$, the last shape-changing event (on average) would have taken place in the last 1 Gy (Figure \ref{fig:nb_col_shape_time}), suggesting that the structure of comet 67P/C-G must have formed in a recent period. 

In Figure \ref{fig:nb_col_form_time}, we plot the average number of events as a function of time for two  potential formation scenarios. In the first scenario, it is assumed that the structure of 67P/C-G formed as a result of a catastrophic break-up of a parent body of $R$ = 3 km. The corresponding number of collisions is then computed from our new $Q^*_D$ scaling. In the second case, we consider impact energies corresponding to a scenario of 67P/C-G formation by low energy sub-catastrophic collisions, as proposed in Paper II. Clearly, the number of events in the later case are substantially larger. This suggests that it may be a more probable formation mechanism than the catastrophic break-up scenario (see a more detailed discussion on this topic in Paper II).

\section{Uncertainties and alternative models}\label{sec:validity}
In this section we discuss some aspects of the  robustness and uncertainties of our modeling approach and alternative models.
\subsection{Critical specific energies}
The values for the  specific catastrophic disruption energies $Q^*_D$ are well defined and follow the expected scaling (Figure  \ref{fig:q_v_disr}). The critical specific impact energies for reshaping are not as well defined and do depend on the material properties. However, we explore a reasonably large range of material properties and also apply large error bars to the results in this case. In any case, there is no doubt that $Q_{reshape} << Q^*_D$ and consequently, there must be many more shape-changing events than catastrophic disruptions. 

\subsection{Dynamical model}

A crucial quantity in the dynamical model is the initial number of comets. The assumption of the existence of $2\times 10^{11}$ comets is in line with estimates of the current Scattered Disk and Oort cloud populations and numerical estimates of the fractions of the primoridal disk that end up in these populations. Both could be wrong, in principle. However the fractions of the primordial disk population implanted in the Scattered Disk and Oort cloud that we use (from \citet{Brasser:2013dw}) are not very different from those found in quite different dynamical models (\citet{Dones:2004dw} for the Oort cloud to Duncan and \citet{Levision:2008lm}, for the Scattered Disk). Therefore, they seem to be robust.

The number of comets used in our model are based also on a flux of Jupiter family comets which is assumed to be currently in a steady state. If this is not the case, the Scattered Disk could be less (or more) populated than predicted by the model. However, we find this unprobable for the following reason. The current estimates for the populations in the Scattered Disk and the Oort cloud are consistent with these two reservoirs being generated from the same parent disk \citep{Brasser:2013dw}. Thus, if the Jupiter family comet flux is now - say 10$\times$ - the mean flux (so to argue for a Scattered Disk 10$\times$ less populated), the same should apply for the flux of long period comets. But the fluxes out of Scattered Disk and Oort cloud follow different processes: for the Scattered Disk, this is resonant diffusion and scattering from Neptune; for the Oort cloud it is stellar perturbations. Therefore, it seems unlikely that both fluxes increased by the same amount relative to the mean values. 

Another crucial quantity in our modeling is the slope of the size distribution $q$, which determines the number of  projectiles of a given size and thus the number of impacts with energies above the critical value. There is an ongoing debate about the form of the size distribution in the Scattered Disk population. We argue that the observations of the crater size distributions in the Pluto system by the New Horizons mission provides one of the best available constraints. The cratering of Pluto and Charon is dominated by the hot population \citep{Greenstreet:2015sg}. All models agree that the hot population and the Scattered Disk population are the same population in terms of physical properties and origin. In fact, the collisional evolution of the hot population is not more severe than that of the Scattered Disk. Both suffered most collisions during the dispersal of the primitive disk (or before, if the dispersal was late). It is true that comets have a shallower distribution \citep{Snodgrass:2011sf} as well as have the craters on the Jovian satellites \citep{Bierhaus:2006,Bierhaus:2009bz}. But this is probably because small comets disintegrate very quickly. On the satellites of Saturn, the crater size distribution is similar to the one expected from a projectile population with a size distribution like that of the main asteroid belt \citep[e.g.][]{Plescia:1982pb,Neukum:2005nw,Neukum:2006nw}, i.e. it is the same as measured by New Horizons on Pluto and Charon.


We note that based on the most recent analysis of the New Horizons data, it has been suggested \citep{Singer:2016sm} that the size distribution for small ($<$ 2 km) objects is shallower ($q \simeq -1.5$) than at large scales. However, this result is still preliminary with uncertainties to be clarified.
As discussed above, the TNO size distribution looks very similar to the size distribution of the asteroid belt, which is a result of a collisional equilibrium (below $\sim$ 100 km).  This suggests that the size distribution of TNOs is in a collisional equilibrium as well. A change of slope below 2 km would produce waves in the TNO size distribution above 2 km. This is not observed, which may indicate that the change of slope is not as pronounced.

To check the effects of a varying slope on our results, we performed additional calculations using $q = -3.3$ for large (> 2 km) and a shallower slope $q_s$ for small (< 2 km) objects. We find that re-shaping collisions could be avoided for $q_s \gtrsim$ -2, which means that if indeed $q_s$ = -1.5, a 67P/C-G-like shape would survive. However, we reiterate that this calculation considers a conservative scenario without any collisional evolution in the primordial disk.

\subsection{Alternative models}
Alternative models to the standard model such as suggested by \citet[][]{Davidsson:2016ds}  predict a much smaller collisional evolution and are  consistent with the idea of comets being primitive unprocessed objects, formed primordially. However, these models require the number of objects in the Scattered Disk to be orders of magnitude smaller. We note that there is no direct observational measure of the Scattered Disk population and all estimates are indirect and pass through models, so such a small number can in principal not be excluded. 

In is not clear, however, how bi-lobe structures would form/survive in these models. 
Previous studies  indicate that the primordial formation of bi-lobed shapes, such as the one of comet 67P/C-G, by direct merging requires extremely low collision velocities of $V/V_{esc}\sim 1$ \citep{Jutzi:2015ja}. This would have to take place at the very early stages of solar system formation, probably while the gas was still present. In the later phases, relative velocities are much higher. In the model of  \citet{Davidsson:2016ds}, average relative velocities are $V$ = 40 m/s during the first 25 Myrs. For kilometer-sized bodies this implies a ratio $V/V_{esc}\sim 40$! In fact, the corresponding specific impact energies are larger than the catastrophic disruption threshold (Figure \ref{fig:q_v_disr_comp}). Our results show that even relative velocities of a few m/s are destructive and lead to reshaping (Figures \ref{fig:qvars}- \ref{fig:qvars300}). Therefore, it is unlikely that primordial bi-lobe structures would survive this phase, and at the same time their formation by collisional merging is implausible due to the high relative velocities.

\section{Summary and conclusions}\label{sec:conclusions}
We have estimated the number of shape-changing collisions for an object with a shape like comet 67P/C-G, considering a dynamical evolution path typical for a Jupiter family comet, using a "standard model" of the early solar system dynamics.

First, we computed the effects of impacts on comet 67P/C-G using a state-of-the-art shock physics code, investigating range of impact conditions and material properties. We found that the shape of comet  67P/C-G, with two lobes connected by a neck, can be destroyed easily, even by impacts with a low specific impact energy.  From these results, scaling laws for the specific energy required for a significant shape alteration ($Q_{reshape}$) were developed. For more general applications, the critical specific energies to alter the shape of generic bi-lobe objects ($Q_{bil}$) was investigated as well. 

These scaling laws for $Q_{reshape}$ and $Q_{bil}$ were then used to analyze the dynamical evolution of a 67P/C-G-like object and generic bi-lobe shapes in terms of shape-changing collisions. We considered a conservative scenario without any collisional evolution before the dynamical instability of the giant planets. Rather, we tracked the collisions during the dispersion of the trans-Neptunian disk caused by the giant planet instability, the formation of a scattered disk of objects and the penetration of tens of objects into the inner solar system. To do this we used a set of simulations \citep{Brasser:2013dw} that produces orbits consistent with the observed JFC population.

We find that even in this conservative scenario, comet 67P/C-G would have experienced a significant number of shape-changing collisions, assuming that its structure formed primordially. For size distributions with $q\le -3.0$, the last reshaping event (on average) would have taken place in the last 1 Gy. The preliminary results of the New Horizons missions concerning the crater size-frequency distribution on Pluto and Charon suggest that the current trans-Neptunian population (i.e. including the Scattered Disk) has a differential power-law size distribution with an exponent $q\simeq-3.3$ \citep{Singer:2015ss}. The possible consequences of a shallower slope for small ($<$ 2 km) objects, as suggested recently by \citet{Singer:2016sm}, are discussed in section \ref{sec:validity}.

It has recently been suggested that rotational fission and reconfiguration may be a dominant structural evolution process for short-period comet nuclei with a two-component structure, provided the volume ratio is larger than $\sim$ 0.2 \citep{Hirabayashi:2016hs}. Our analysis of impacts on generic bi-lobe shapes shows that they would have experienced a substantial number of  collisions with energies sufficient to destroy their two-component structure. This strongly suggests that the two-component body which is required to exist at the beginning of the fission-merging cycle cannot be primordial. 

Thus, according to our model, comets are not primordial in the sense that their current shape and structure did not form in the initial stages of the formation of the solar system. Rather, they evolve through the effect of collisions and the final shape is a result of the last major reshaping impact, possibly within the last 1 Gy. A scenario of a late formation of  67P/C-G-like two-component structures is presented in Paper II.

It is clear that the results presented here are based on the assumption that the standard model of dynamical evolution is correct. Although some of its parameters are debated, as discussed in section \ref{sec:validity}, we believe that the model is robust. We note that it is so far the only model which produces the correct number of objects in the inner solar system with orbits consistent with the observed JFC population.

Our results clearly show that if this standard model of solar system dynamics is correct, it means that the cometary nuclei as they are observed today must be collisionally processed objects. Therefore, the remaining important question is whether or not such collisionally processed bodies can still have primitive properties (i.e. high porosity, presence of supervolatiles). If this is not the case, then the standard model must be wrong. This would mean for instance that either the primordial disk was dynamically cold and contained a much lower number of objects, as proposed by \citet[][]{Davidsson:2016ds} or that there is a lack of small comets, implying an abrupt change in the slope of the size distribution.

However, the analysis of the outcomes of the detailed impact modeling carried out here (for shape-changing impacts and catastrophic disruptions) suggest that collisionally processed cometary nuclei can still have a high porosity, and could have retained their volatiles, since there is no significant large-scale heating. Therefore, they may still look primitive, meaning that the standard model is consistent with the observations of comet 67P/C-G. This question is investigated further in Paper II and also in an ongoing study of bi-lobe formation in large-scale catastrophic disruptions (Schwartz et al., 2016, in prep).

Primordial or not, the structure of comet 67P/C-G is an important probe of the dynamical history of small bodies.

\begin{acknowledgements}
     M.J. and W.B. acknowledge support from the Swiss NCCR PlanetS. A.T. wishes to thank OCA for their kind hospitality during her stay there. We thank the referees B. Davidsson and J. A. Fernandez for their thorough review which helped improve the paper substantially.
 \end{acknowledgements}

%
\bibliographystyle{aa} 
\bibliography{../bibdata.bib} 

\begin{thebibliography}{63}
\expandafter\ifx\csname natexlab\endcsname\relax\def\natexlab#1{#1}\fi

\bibitem[{Asphaug \& Benz(1994)}]{Asphaug:1994cya}
Asphaug, E. \& Benz, W. 1994, Nature, 370, 120

\bibitem[{Benz \& Asphaug(1995)}]{Benz:1995hx}
Benz, W. \& Asphaug, E. 1995, Computer Physics Communications, 87, 253

\bibitem[{Benz \& Asphaug(1999)}]{Benz:1999cj}
Benz, W. \& Asphaug, E. 1999, ICARUS, 142, 5

\bibitem[{Biele {et~al.}(2015)Biele, Ulamec, Maibaum, Roll, Witte, Jurado,
  Mu{\~n}oz, Arnold, Auster, Casas, Faber, Fantinati, Finke, Fischer, Geurts,
  G{\"u}ttler, Heinisch, H{\'e}rique, Hviid, Kargl, Knapmeyer, Knollenberg,
  Kofman, K{\"o}mle, K{\"u}hrt, Lommatsch, Mottola, Pardo~de Santayana,
  Remetean, Scholten, Seidensticker, Sierks, \& Spohn}]{biele:2015bu}
Biele, J., Ulamec, S., Maibaum, M., {et~al.} 2015, Science, 349, 9816

\bibitem[{{Bierhaus}(2006)}]{Bierhaus:2006}
{Bierhaus}, E.~B. 2006, in LPI Contributions, Vol. 1320, Workshop on Surface
  Ages and Histories: Issues in Planetary Chronology, 14--15

\bibitem[{{Bierhaus} {et~al.}(2009){Bierhaus}, {Zahnle}, \&
  {Chapman}}]{Bierhaus:2009bz}
{Bierhaus}, E.~B., {Zahnle}, K., \& {Chapman}, C.~R. 2009, {Europa's Crater
  Distributions and Surface Ages}, ed. R.~T. {Pappalardo}, W.~B. {McKinnon}, \&
  K.~K. {Khurana}, 161

\bibitem[{Boehnhardt(2004)}]{Boehnhardt:2004uu}
Boehnhardt, H. 2004, Comets II, University of Arizona Press., 301

\bibitem[{Brasser \& Morbidelli(2013)}]{Brasser:2013dw}
Brasser, R. \& Morbidelli, A. 2013, Icarus, 225, 40

\bibitem[{{Capaccioni} {et~al.}(2015){Capaccioni}, {Coradini}, {Filacchione},
  {Erard}, {Arnold}, {Drossart}, {De Sanctis}, {Bockelee-Morvan}, {Capria},
  {Tosi}, {Leyrat}, {Schmitt}, {Quirico}, {Cerroni}, {Mennella}, {Raponi},
  {Ciarniello}, {McCord}, {Moroz}, {Palomba}, {Ammannito}, {Barucci},
  {Bellucci}, {Benkhoff}, {Bibring}, {Blanco}, {Blecka}, {Carlson}, {Carsenty},
  {Colangeli}, {Combes}, {Combi}, {Crovisier}, {Encrenaz}, {Federico}, {Fink},
  {Fonti}, {Ip}, {Irwin}, {Jaumann}, {Kuehrt}, {Langevin}, {Magni}, {Mottola},
  {Orofino}, {Palumbo}, {Piccioni}, {Schade}, {Taylor}, {Tiphene}, {Tozzi},
  {Beck}, {Biver}, {Bonal}, {Combe}, {Despan}, {Flamini}, {Fornasier},
  {Frigeri}, {Grassi}, {Gudipati}, {Longobardo}, {Markus}, {Merlin}, {Orosei},
  {Rinaldi}, {Stephan}, {Cartacci}, {Cicchetti}, {Giuppi}, {Hello}, {Henry},
  {Jacquinod}, {Noschese}, {Peter}, {Politi}, {Reess}, \&
  {Semery}}]{Capaccioni:2015cc}
{Capaccioni}, F., {Coradini}, A., {Filacchione}, G., {et~al.} 2015, Science,
  347, aaa0628

\bibitem[{Cuzzi {et~al.}(2010)Cuzzi, Hogan, \& Bottke}]{Cuzzi:2010iv}
Cuzzi, J.~N., Hogan, R.~C., \& Bottke, W.~F. 2010, ICARUS, 208, 518

\bibitem[{{Davidsson} {et~al.}(2016){Davidsson}, {Sierks}, {G{\"u}ttler},
  {Marzari}, {Pajola}, {Rickman}, {A'Hearn}, {Auger}, {El-Maarry}, {Fornasier},
  {Guti{\'e}rrez}, {Keller}, {Massironi}, {Snodgrass}, {Vincent}, {Barbieri},
  {Lamy}, {Rodrigo}, {Koschny}, {Barucci}, {Bertaux}, {Bertini}, {Cremonese},
  {Da Deppo}, {Debei}, {De Cecco}, {Feller}, {Fulle}, {Groussin}, {Hviid},
  {H{\"o}fner}, {Ip}, {Jorda}, {Knollenberg}, {Kovacs}, {Kramm}, {K{\"u}hrt},
  {K{\"u}ppers}, {La Forgia}, {Lara}, {Lazzarin}, {Lopez Moreno},
  {Moissl-Fraund}, {Mottola}, {Naletto}, {Oklay}, {Thomas}, \&
  {Tubiana}}]{Davidsson:2016ds}
{Davidsson}, B.~J.~R., {Sierks}, H., {G{\"u}ttler}, C., {et~al.} 2016, \aap,
  592, A63

\bibitem[{{Dones} {et~al.}(2004){Dones}, {Weissman}, {Levison}, \&
  {Duncan}}]{Dones:2004dw}
{Dones}, L., {Weissman}, P.~R., {Levison}, H.~F., \& {Duncan}, M.~J. 2004,
  {Oort cloud formation and dynamics}, ed. M.~C. {Festou}, H.~U. {Keller}, \&
  H.~A. {Weaver}, 153--174

\bibitem[{{Duncan} \& {Levison}(1997)}]{Duncan:1997dl}
{Duncan}, M.~J. \& {Levison}, H.~F. 1997, Science, 276, 1670

\bibitem[{{Fujiwara} {et~al.}(2006){Fujiwara}, {Kawaguchi}, {Yeomans}, {Abe},
  {Mukai}, {Okada}, {Saito}, {Yano}, {Yoshikawa}, {Scheeres}, {Barnouin-Jha},
  {Cheng}, {Demura}, {Gaskell}, {Hirata}, {Ikeda}, {Kominato}, {Miyamoto},
  {Nakamura}, {Nakamura}, {Sasaki}, \& {Uesugi}}]{Fujiwara:2006fk}
{Fujiwara}, A., {Kawaguchi}, J., {Yeomans}, D.~K., {et~al.} 2006, Science, 312,
  1330

\bibitem[{{Gomes} {et~al.}(2005){Gomes}, {Levison}, {Tsiganis}, \&
  {Morbidelli}}]{Gomes:2005gl}
{Gomes}, R., {Levison}, H.~F., {Tsiganis}, K., \& {Morbidelli}, A. 2005, \nat,
  435, 466

\bibitem[{{Greenstreet} {et~al.}(2015){Greenstreet}, {Gladman}, \&
  {McKinnon}}]{Greenstreet:2015sg}
{Greenstreet}, S., {Gladman}, B., \& {McKinnon}, W.~B. 2015, \icarus, 258, 267

\bibitem[{{G{\"u}ttler} {et~al.}(2009){G{\"u}ttler}, {Krause}, {Geretshauser},
  {Speith}, \& {Blum}}]{Guettler:2009gk}
{G{\"u}ttler}, C., {Krause}, M., {Geretshauser}, R.~J., {Speith}, R., \&
  {Blum}, J. 2009, \apj, 701, 130

\bibitem[{{H{\"a}ssig} {et~al.}(2015){H{\"a}ssig}, {Altwegg}, {Balsiger}, \&
  {etal.}}]{Haessig:2015he}
{H{\"a}ssig}, M., {Altwegg}, K., {Balsiger}, H., \& {etal.} 2015, Science, 347

\bibitem[{{Hirabayashi} {et~al.}(2016){Hirabayashi}, {Scheeres}, {Chesley},
  {Marchi}, {McMahon}, {Steckloff}, {Mottola}, {Naidu}, \&
  {Bowling}}]{Hirabayashi:2016hs}
{Hirabayashi}, M., {Scheeres}, D.~J., {Chesley}, S.~R., {et~al.} 2016, \nat,
  534, 352

\bibitem[{Housen \& Holsapple(1990)}]{Housen:1990hh}
Housen, K.~R. \& Holsapple, K.~A. 1990, Icarus (ISSN 0019-1035), 84, 226

\bibitem[{Housen \& Holsapple(2003)}]{Housen:2003bz}
Housen, K.~R. \& Holsapple, K.~A. 2003, Icarus, 163, 102

\bibitem[{{Johansen} {et~al.}(2007){Johansen}, {Oishi}, {Mac Low}, {Klahr},
  {Henning}, \& {Youdin}}]{Johansen:2007jo}
{Johansen}, A., {Oishi}, J.~S., {Mac Low}, M.-M., {et~al.} 2007, \nat, 448,
  1022

\bibitem[{Jutzi(2015)}]{Jutzi:2015gb}
Jutzi, M. 2015, Planetary and Space Science, 107, 3

\bibitem[{Jutzi \& Asphaug(2015)}]{Jutzi:2015ja}
Jutzi, M. \& Asphaug, E. 2015, Science, 348, 1

\bibitem[{Jutzi {et~al.}(2008)Jutzi, Benz, \& Michel}]{Jutzi:2008kp}
Jutzi, M., Benz, W., \& Michel, P. 2008, Icarus, 198, 242

\bibitem[{Jutzi {et~al.}(2015)Jutzi, Holsapple, W{\"u}nnemann, \&
  Michel}]{Jutzi:2015ux}
Jutzi, M., Holsapple, K.~A., W{\"u}nnemann, K., \& Michel, P. 2015, Asteroids
  IV, 1

\bibitem[{Jutzi {et~al.}(2010)Jutzi, Michel, Benz, \&
  Richardson}]{Jutzi:2010bf}
Jutzi, M., Michel, P., Benz, W., \& Richardson, D.~C. 2010, Icarus, 207, 54

\bibitem[{{Kaib} \& {Chambers}(2016)}]{Kaib:2016kc}
{Kaib}, N.~A. \& {Chambers}, J.~E. 2016, \mnras, 455, 3561

\bibitem[{{Kataoka} {et~al.}(2013){Kataoka}, {Tanaka}, {Okuzumi}, \&
  {Wada}}]{Kataoka:2013kt}
{Kataoka}, A., {Tanaka}, H., {Okuzumi}, S., \& {Wada}, K. 2013, \aap, 557, L4

\bibitem[{{Klinger}(1981)}]{Klinger:1981}
{Klinger}, J. 1981, \icarus, 47, 320

\bibitem[{{Kofman} {et~al.}(2015){Kofman}, {Herique}, {Barbin}, {Barriot},
  {Ciarletti}, {Clifford}, {Edenhofer}, {Elachi}, {Eyraud}, {Goutail}, {Heggy},
  {Jorda}, {Lasue}, {Levasseur-Regourd}, {Nielsen}, {Pasquero}, {Preusker},
  {Puget}, {Plettemeier}, {Rogez}, {Sierks}, {Statz}, {Svedhem}, {Williams},
  {Zine}, \& {Van Zyl}}]{Kofman:2015kh}
{Kofman}, W., {Herique}, A., {Barbin}, Y., {et~al.} 2015, Science, 349

\bibitem[{Leinhardt \& Stewart(2009)}]{Leinhardt:2009ic}
Leinhardt, Z.~M. \& Stewart, S.~T. 2009, Icarus, 199, 542

\bibitem[{Leinhardt \& Stewart(2012)}]{Leinhardt:2012ls}
Leinhardt, Z.~M. \& Stewart, S.~T. 2012, The Astrophysical Journal, 745, 79

\bibitem[{{Levison} {et~al.}(2008){Levison}, {Morbidelli}, {Van Laerhoven},
  {Gomes}, \& {Tsiganis}}]{Levision:2008lm}
{Levison}, H.~F., {Morbidelli}, A., {Van Laerhoven}, C., {Gomes}, R., \&
  {Tsiganis}, K. 2008, \icarus, 196, 258

\bibitem[{{Lorek} {et~al.}(2016){Lorek}, {Gundlach}, {Lacerda}, \&
  {Blum}}]{Lorek:2016lg}
{Lorek}, S., {Gundlach}, B., {Lacerda}, P., \& {Blum}, J. 2016, \aap, 587, A128

\bibitem[{{Massironi} {et~al.}(2015){Massironi}, {Simioni}, {Marzari},
  {Cremonese}, {Giacomini}, {Pajola}, {Jorda}, {Naletto}, {Lowry}, {El-Maarry},
  {Preusker}, {Scholten}, {Sierks}, {Barbieri}, {Lamy}, {Rodrigo}, {Koschny},
  {Rickman}, {Keller}, {A'Hearn}, {Agarwal}, {Auger}, {Barucci}, {Bertaux},
  {Bertini}, {Besse}, {Bodewits}, {Capanna}, {da Deppo}, {Davidsson}, {Debei},
  {de Cecco}, {Ferri}, {Fornasier}, {Fulle}, {Gaskell}, {Groussin},
  {Guti{\'e}rrez}, {G{\"u}ttler}, {Hviid}, {Ip}, {Knollenberg}, {Kovacs},
  {Kramm}, {K{\"u}hrt}, {K{\"u}ppers}, {La Forgia}, {Lara}, {Lazzarin}, {Lin},
  {Lopez Moreno}, {Magrin}, {Michalik}, {Mottola}, {Oklay}, {Pommerol},
  {Thomas}, {Tubiana}, \& {Vincent}}]{Massironi:2015ma}
{Massironi}, M., {Simioni}, E., {Marzari}, F., {et~al.} 2015, \nat, 526, 402

\bibitem[{{Melosh}(1989)}]{Melosh:1989}
{Melosh}, H.~J. 1989, {Impact cratering: A geologic process}

\bibitem[{Morbidelli {et~al.}(2009)Morbidelli, Bottke, Nesvorn{\'y}, \&
  Levison}]{Morbidelli:2009dd}
Morbidelli, A., Bottke, W.~F., Nesvorn{\'y}, D., \& Levison, H.~F. 2009,
  ICARUS, 204, 558

\bibitem[{{Morbidelli} {et~al.}(2012){Morbidelli}, {Marchi}, {Bottke}, \&
  {Kring}}]{Morbidelli:2012ms}
{Morbidelli}, A., {Marchi}, S., {Bottke}, W.~F., \& {Kring}, D.~A. 2012, Earth
  and Planetary Science Letters, 355, 144

\bibitem[{{Morbidelli} \& {Rickman}(2015)}]{Morbidelli:2015vm}
{Morbidelli}, A. \& {Rickman}, H. 2015, \aap, 583, A43

\bibitem[{Mumma {et~al.}(1993)Mumma, Weissman, \& Stern}]{Mumma:1993mw}
Mumma, M.~J., Weissman, P.~R., \& Stern, S.~A. 1993, In: Protostars and planets
  III (A93-42937 17-90), 1177

\bibitem[{{Neukum} {et~al.}(2006){Neukum}, {Wagner}, {Wolf}, \&
  {Denk}}]{Neukum:2006nw}
{Neukum}, G., {Wagner}, R., {Wolf}, U., \& {Denk}, T. 2006, in European
  Planetary Science Congress 2006, 610

\bibitem[{{Neukum} {et~al.}(2005){Neukum}, {Wagner}, {Denk}, {Porco}, \&
  {Cassini Iss Team}}]{Neukum:2005nw}
{Neukum}, G., {Wagner}, R.~J., {Denk}, T., {Porco}, C.~C., \& {Cassini Iss
  Team}. 2005, in Lunar and Planetary Inst.~Technical Report, Vol.~36, 36th
  Annual Lunar and Planetary Science Conference, ed. S.~{Mackwell} \&
  E.~{Stansbery}

\bibitem[{Nyffeler(2004)}]{Nyffeler:2004tz}
Nyffeler, B. 2004, PhD thesis, University of Bern

\bibitem[{{P{\"a}tzold} {et~al.}(2016){P{\"a}tzold}, {Andert}, {Hahn}, {Asmar},
  {Barriot}, {Bird}, {H{\"a}usler}, {Peter}, {Tellmann}, {Gr{\"u}n},
  {Weissman}, {Sierks}, {Jorda}, {Gaskell}, {Preusker}, \&
  {Scholten}}]{Paetzold:2016pa}
{P{\"a}tzold}, M., {Andert}, T., {Hahn}, M., {et~al.} 2016, \nat, 530, 63

\bibitem[{{Plescia} \& {Boyce}(1982)}]{Plescia:1982pb}
{Plescia}, J.~B. \& {Boyce}, J.~M. 1982, \nat, 295, 285

\bibitem[{{Rickman} {et~al.}(2015){Rickman}, {Marchi}, {A'Hearn}, {Barbieri},
  {El-Maarry}, {G{\"u}ttler}, {Ip}, {Keller}, {Lamy}, {Marzari}, {Massironi},
  {Naletto}, {Pajola}, {Sierks}, {Koschny}, {Rodrigo}, {Barucci}, {Bertaux},
  {Bertini}, {Cremonese}, {Da Deppo}, {Debei}, {De Cecco}, {Fornasier},
  {Fulle}, {Groussin}, {Guti{\'e}rrez}, {Hviid}, {Jorda}, {Knollenberg},
  {Kramm}, {K{\"u}hrt}, {K{\"u}ppers}, {Lara}, {Lazzarin}, {Lopez Moreno},
  {Michalik}, {Sabau}, {Thomas}, {Vincent}, \& {Wenzel}}]{Rickman:2015wu}
{Rickman}, H., {Marchi}, S., {A'Hearn}, M.~F., {et~al.} 2015, \aap, 583, A44

\bibitem[{{Robie} \& Hemingway(1982)}]{Robie:1982rh}
{Robie}, A. \& Hemingway, B. 1982, American Mineralogist, 67, 470

\bibitem[{{Rotundi} {et~al.}(2015){Rotundi}, {Sierks}, {Della Corte}, {Fulle},
  {Gutierrez}, {Lara}, {Barbieri}, {Lamy}, {Rodrigo}, {Koschny}, {Rickman},
  {Keller}, {L{\'o}pez-Moreno}, {Accolla}, {Agarwal}, {A'Hearn}, {Altobelli},
  {Angrilli}, {Barucci}, {Bertaux}, {Bertini}, {Bodewits}, {Bussoletti},
  {Colangeli}, {Cosi}, {Cremonese}, {Crifo}, {Da Deppo}, {Davidsson}, {Debei},
  {De Cecco}, {Esposito}, {Ferrari}, {Fornasier}, {Giovane}, {Gustafson},
  {Green}, {Groussin}, {Gr{\"u}n}, {G{\"u}ttler}, {Herranz}, {Hviid}, {Ip},
  {Ivanovski}, {Jer{\'o}nimo}, {Jorda}, {Knollenberg}, {Kramm}, {K{\"u}hrt},
  {K{\"u}ppers}, {Lazzarin}, {Leese}, {L{\'o}pez-Jim{\'e}nez}, {Lucarelli},
  {Lowry}, {Marzari}, {Epifani}, {McDonnell}, {Mennella}, {Michalik}, {Molina},
  {Morales}, {Moreno}, {Mottola}, {Naletto}, {Oklay}, {Ortiz}, {Palomba},
  {Palumbo}, {Perrin}, {Rodr{\'{\i}}guez}, {Sabau}, {Snodgrass}, {Sordini},
  {Thomas}, {Tubiana}, {Vincent}, {Weissman}, {Wenzel}, {Zakharov}, \&
  {Zarnecki}}]{Rotundi:2015rs}
{Rotundi}, A., {Sierks}, H., {Della Corte}, V., {et~al.} 2015, Science, 347,
  aaa3905

\bibitem[{{Sierks} {et~al.}(2015){Sierks}, {Barbieri}, {Lamy}, {Rodrigo},
  {Koschny}, {Rickman}, {Keller}, {Agarwal}, {A'Hearn}, {Angrilli}, {Auger},
  {Barucci}, {Bertaux}, {Bertini}, {Besse}, {Bodewits}, {Capanna}, {Cremonese},
  {Da Deppo}, {Davidsson}, {Debei}, {De Cecco}, {Ferri}, {Fornasier}, {Fulle},
  {Gaskell}, {Giacomini}, {Groussin}, {Gutierrez-Marques}, {Guti{\'e}rrez},
  {G{\"u}ttler}, {Hoekzema}, {Hviid}, {Ip}, {Jorda}, {Knollenberg}, {Kovacs},
  {Kramm}, {K{\"u}hrt}, {K{\"u}ppers}, {La Forgia}, {Lara}, {Lazzarin},
  {Leyrat}, {Lopez Moreno}, {Magrin}, {Marchi}, {Marzari}, {Massironi},
  {Michalik}, {Moissl}, {Mottola}, {Naletto}, {Oklay}, {Pajola}, {Pertile},
  {Preusker}, {Sabau}, {Scholten}, {Snodgrass}, {Thomas}, {Tubiana}, {Vincent},
  {Wenzel}, {Zaccariotto}, \& {P{\"a}tzold}}]{Sierks:2015sb}
{Sierks}, H., {Barbieri}, C., {Lamy}, P.~L., {et~al.} 2015, Science, 347,
  aaa1044

\bibitem[{{Singer} {et~al.}(2016){Singer}, {McKinnon}, {Greenstreet},
  {Gladman}, {Parker}, {Robbins}, {Schenk}, {Stern}, {Bray}, {Spencer},
  {Weaver}, {Beyer}, {Young}, {Moore}, {Olkin}, {Ennico}, {Binzel}, {Grundy},
  \& {New Horizons Geology Geophysics and Imaging Science Theme
  Team}}]{Singer:2016sm}
{Singer}, K.~N., {McKinnon}, W.~B., {Greenstreet}, S., {et~al.} 2016, in
  AAS/Division for Planetary Sciences Meeting Abstracts, Vol.~48, AAS/Division
  for Planetary Sciences Meeting Abstracts

\bibitem[{{Singer} {et~al.}(2015){Singer}, {Schenk}, {Robbins}, {Bray},
  {McKinnon}, {Moore}, {Spencer}, {Stern}, {Grundy}, {Howett}, {Dalle Ore},
  {Beyer}, {Parker}, {Porter}, {Zangari}, {Young}, {Olkin}, \&
  {Ennico}}]{Singer:2015ss}
{Singer}, K.~N., {Schenk}, P.~M., {Robbins}, S.~J., {et~al.} 2015, in
  AAS/Division for Planetary Sciences Meeting Abstracts, Vol.~47, AAS/Division
  for Planetary Sciences Meeting Abstracts, 102.02

\bibitem[{Skorov \& Blum(2012)}]{Skorov:2012ks}
Skorov, Y. \& Blum, J. 2012, ICARUS, 1

\bibitem[{{Snodgrass} {et~al.}(2011){Snodgrass}, {Fitzsimmons}, {Lowry}, \&
  {Weissman}}]{Snodgrass:2011sf}
{Snodgrass}, C., {Fitzsimmons}, A., {Lowry}, S.~C., \& {Weissman}, P. 2011,
  \mnras, 414, 458

\bibitem[{Speith(2006)}]{Speith:2006}
Speith, R. 2006, Habilitation, University of T\"ubingen

\bibitem[{{Steckloff} {et~al.}(2015){Steckloff}, {Johnson}, {Bowling}, {Jay
  Melosh}, {Minton}, {Lisse}, \& {Battams}}]{Steckloff:2015sj}
{Steckloff}, J.~K., {Johnson}, B.~C., {Bowling}, T., {et~al.} 2015, \icarus,
  258, 430

\bibitem[{{Toliou} {et~al.}(2016){Toliou}, {Morbidelli}, \&
  {Tsiganis}}]{Toliou:2016tm}
{Toliou}, A., {Morbidelli}, A., \& {Tsiganis}, K. 2016, \aap, 592, A72

\bibitem[{{Tsiganis} {et~al.}(2005){Tsiganis}, {Gomes}, {Morbidelli}, \&
  {Levison}}]{Tsiganis:2005tg}
{Tsiganis}, K., {Gomes}, R., {Morbidelli}, A., \& {Levison}, H.~F. 2005, \nat,
  435, 459

\bibitem[{Weidenschilling(1997)}]{Weidenschilling:1997im}
Weidenschilling, S.~J. 1997, ICARUS, 127, 290

\bibitem[{Weissman {et~al.}(2004)Weissman, Asphaug, \&
  Lowry}]{Weissmann:2004wa}
Weissman, P.~R., Asphaug, E., \& Lowry, S.~C. 2004, Comets II, 337

\bibitem[{{Windmark} {et~al.}(2012{\natexlab{a}}){Windmark}, {Birnstiel},
  {G{\"u}ttler}, {Blum}, {Dullemond}, \& {Henning}}]{Windmark:2012wbg}
{Windmark}, F., {Birnstiel}, T., {G{\"u}ttler}, C., {et~al.}
  2012{\natexlab{a}}, \aap, 540, A73

\bibitem[{{Windmark} {et~al.}(2012{\natexlab{b}}){Windmark}, {Birnstiel},
  {Ormel}, \& {Dullemond}}]{Windmark:2012wbo}
{Windmark}, F., {Birnstiel}, T., {Ormel}, C.~W., \& {Dullemond}, C.~P.
  2012{\natexlab{b}}, \aap, 544, L16

\bibitem[{{Youdin} \& {Goodman}(2005)}]{Youdin:2005yg}
{Youdin}, A.~N. \& {Goodman}, J. 2005, \apj, 620, 459

\end{thebibliography}
%

\end{document}